\documentclass[12pt]{article}
\usepackage{amsmath}
\usepackage{amssymb}
\usepackage{jheppub}
\usepackage{amsmath, amsthm, amsfonts, amssymb, bbm}
\usepackage[mathscr]{eucal}
\usepackage{graphics}
\usepackage{xypic}
\usepackage[all]{xy}\xyoption{rotate}
\usepackage[english]{babel}
\usepackage[font=small,format=plain,labelfont=bf,up]{caption}
\usepackage[bbgreekl]{mathbbol}
\usepackage{tikz}

\DeclareMathOperator{\Sym}{Sym}
\newcommand{\vac}{\left|\emptyset\right\rangle}
\newcommand{\be}{\begin{equation}}
\newcommand{\ee}{\end{equation}}

\newcommand{\eps}{\varepsilon}
\newcommand{\bC}{\mathbb{C}}
\newcommand{\bZ}{\mathbb{Z}}
\theoremstyle{definition}
\theoremstyle{plain}

\theoremstyle{definition}
\theoremstyle{plain}

\theoremstyle{definition}
\theoremstyle{definition}
\theoremstyle{definition}
\theoremstyle{definition}
\theoremstyle{definition}    
    
\theoremstyle{definition}

\DeclareMathOperator{\Hom}{Hom}
\DeclareMathOperator{\Ext}{Ext}
\DeclareMathOperator{\modu}{mod}
\DeclareMathOperator{\Fc}{Fc}
\DeclareMathOperator{\Ff}{Ff}
\DeclareMathOperator{\Sb}{Sb}
\DeclareMathOperator{\Vir}{Vir}
\newcommand{\psu}{\mathfrak{psu}}
\newcommand{\cV}{\mathcal{V}}
\newcommand{\cE}{\mathcal{E}}
\newcommand{\cA}{\mathcal{A}}
\newcommand{\cP}{\mathcal{P}}
\newcommand{\cO}{\mathcal{O}}
\newcommand{\cW}{\mathcal{W}}
\newcommand{\ccH}{\mathcal{H}}

\newcommand{\C}{{\mathbb C}}
\renewcommand{\b}{\sf b}
\renewcommand{\c}{\sf c}
\newcommand{\into}{\hookrightarrow}

\renewcommand{\d}{\mathrm{d}}
\newcommand{\g}{\mathfrak{g}}

\newcommand{\op}{\operatorname}
\newcommand{\mc}{\mathcal}
 
\newcommand{\mf}{\mathfrak}
\newcommand{\what}{\widehat}

\newcommand{\catfock}{\mathcal C^{\mathcal F}}
\newcommand{\catfocklog}{\mathcal C^{\mathcal F}_{\text{log}}}
\newcommand{\cat}{\mathcal C}
\newcommand{\catlog}{\mathcal C_{\text{log}}}
\newcommand{\catA}{\mathcal C^A_{\text{loc}}}

\newcommand{\catB}{\mathcal C^B_{\text{loc}}}

\newtheorem*{conjecture}{Conjecture}

\title{Higgs and Coulomb branches from vertex operator algebras}
\author[1]{Kevin Costello,}
\author[2]{Thomas Creutzig,}
\author[1]{Davide Gaiotto}
\affiliation[1]{Perimeter Institute for Theoretical Physics, Waterloo, Ontario, Canada N2L 2Y5}
\affiliation[2]{University of Alberta, Edmonton, AB T6G 2G1}
\abstract{We formulate a conjectural relation between the category of line defects in topologically twisted 3d ${\cal N}=4$ supersymmetric quantum field theories
and categories of modules for Vertex Operator Algebras of boundary local operators for the theories. We test the conjecture in several examples and 
provide some partial proofs for standard classes of gauge theories.}

\begin{document}
\maketitle

\section{Introduction}
There are various intertwined relations between supersymmetric gauge theories and Vertex Operator Algebras 
\cite{Nekrasov:2002qd,Alday:2009aq,Nekrasov:2010ka,Beem:2013sza,Beem:2014rza,Beem:2014kka,Bonetti:2018fqz,Gaiotto:2016wcv,Gaiotto:2017euk,CG17,Costello:2018fnz}.
In many of these constructions the VOA emerges as the local operator algebra of some QFT which is topological away from some special two-dimensional location or defect 
and holomorphic at the defect. Holomorphicity guarantees that the local operators at that location will have meromorphic OPE's and form a vertex algebra. 
Topological invariance away from the two-dimensional locus guarantees the existence of a two-dimensional stress tensor among these local operators. 

A prototypical example is that of a 3d TFT with a holomorphic boundary condition. In a physical context, we are familiar to holomorphic boundary conditions for Chern-Simons gauge theory, say supporting WZW vertex algebras of local operators. In this paper we concern ourselves with analogous configurations involving topological twists of three-dimensional gauge theories.

In the physical context, there is a close relation between the properties of the bulk TFT $T_{3d}$ and of the boundary VOA $V$. The boundary VOA is generically rational, 
with finite-dimensional spaces of conformal blocks and a modular tensor category of VOA modules. That data essentially {\it defines} a 3d TFT $T[V]$. The $T[V]$ TFT 
does not have to be the same as $T_{3d}$, but the two are closely related. In particular, we can map each line defect $\ell$ of $T_{3d}$ to the $V$-module $M_\ell$ 
of local operators at the location where $\ell$ ends on the boundary. Topological local operators interpolating between two line defects 
will map to morphisms between the corresponding modules. This functor will be compatible with operations of fusion, braiding, sewing, etc. Similarly, each state of $T_{3d}$
on some Riemann surface will map to a conformal block of $V$ on the same surface, in a manner compatible with the action of the mapping class group of the surface. 
In sufficiently non-degenerate situations, $V$ will uniquely determine $T_{3d}$.

The theories relevant to this paper, topological twists of 3d ${\cal N}=4$ gauge theories, are TFTs of cohomological type, which have looser properties than 
physical TFTs. In particular, they lack unitarity and various finiteness constraints. Their boundary conditions support VOAs 
which can be not rational and have intricate categories of modules admitting non-trivial extensions. Furthermore, the relations between 
bulk line defects and vertex algebra modules may hold in some differential-graded, or derived sense. 

The precise mathematical definition of the category of line defects in twisted 3d ${\cal N}=4$ gauge theories is not yet fully understood, 
but is being actively investigate in light of important connections to Symplectic Duality and Geometric Langlands programs. In this paper we will assume some 
basic properties of such a category. For example, we expect the bulk local operators, i.e. the endomorphisms of the trivial line, to coincide with the algebra of functions on the 
Higgs or Coulomb branches of vacua of the theory, depending on the choice of topological twist. The Higgs branch of a standard gauge theory is 
easily computed as a classical symplectic quotient, but the Coulomb branch receives difficult quantum corrections \cite{Seiberg:1996bs,Seiberg:1996nz,deBoer:1996ck,Kapustin:1999ha,Hanany:1996ie,Gaiotto:2008ak,Bullimore:2015lsa}. A sharp mathematical proposal was recently given in \cite{Nakajima:2015txa,Braverman:2016wma}.

Holomorphic boundary conditions for twisted 3d ${\cal N}=4$ gauge theories were recently proposed by two of the authors \cite{Costello:2018fnz}. 
The objective of this paper is to study the relation between the bulk topological data and the properties of modules for the corresponding boundary VOAs.
In particular, we would like to compare the algebra of functions on the Higgs or Coulomb branches with the algebra of derived endomorphisms of the vacuum module 
for the boundary VOAs. The analysis of the most basic examples will immediately show us the importance of the ``derived'' part of this statement. In turns, that will present us with an additional challenge: in order to define or compute derived endomorphisms we will need to identify some ``good'' category of modules for the boundary VOAs,
which may or may not coincide with the categories of modules which are commonly studied in the VOA literature.

In this paper we will not give a full solution of these challenges. Instead, our work will be of a somewhat experimental nature: we will study 
increasingly complicated examples and identify which choices lead to a match between the VOA and TFT answers. We hope that our work will motivate 
further research in supersymmetric/topological QFTs, VOAs and Geometric Representation theory which will allow a sharp formulation and a proof of the rough conjecture:
\begin{itemize}
\item {\it The category of line defects for a twisted 3d ${\cal N}=4$ gauge theory can be identified with a derived category of modules for some boundary VOA.}
\end{itemize}

Such a relation can be used both ways. On one hand, it will explain, organizes and predicts non-trivial properties of 
important classes of VOA's, endowing them with some generalized notion of rationality. On the other hand, 
the VOAs themselves can be effective computational tools to study the bulk TFTs.  

One final observation is that the algebras of endomorphisms of line defects in twisted 3d ${\cal N}=4$ gauge theories admit interesting 
``quantum deformations'' associated to $\Omega$ deformations of the theory \cite{Yagi:2014toa}. We expect these quantum deformations to also arise 
from VOA constructions, perhaps working equivariantly for loop rotations. 

\subsection{Structure of the paper}
In Section \ref{sec:hyper} we will discuss at length, from different perspectives, the most basic example: the two 
twists of the free hypermultiplet SQFT. In Section \ref{sec:hypertwo} we will discuss SQED$_1$ and the mirror symmetry relation to 
a free hypermultiplet. In Section \ref{sec:tsutwo} we will discuss in detail the next simplest Abelian gauge theory, SQED$_2$. In Section \ref{sec:abelian} we will discuss more general Abelian gauge theories. In Section \ref{sec:higgs} we will sketch a general argument for the relation 
between the VOA at Neumann boundary conditions and the algebra of functions on the Coulomb branch of all standard gauge theories. In Section \ref{sec:coulomb} we will sketch a general argument for the relation 
between the VOA at Dirichlet boundary conditions and the algebra of functions on the Higgs branch of all standard gauge theories. In Section \ref{sec:ad} we will discuss some VOAs which are conjecturally related to Argyres-Douglas-type theories, which do not admit a standard gauge theory Lagrangian description. Section \ref{sec:voa} contains direct VOA calculations of extensions of modules and conjectural VOA descriptions of the associated module categories. 
\section{A basic example: the free hypermultiplet} \label{sec:hyper}

The boundary VOAs for the theory of a free hypermultiplet are among the simplest examples of 
``logarithmic'' VOAs with representations that are indecomposable but not completely reducible \cite{Creutzig:2013hma}: the ``symplectic bosons'' VOA $\mathrm{Sb}$ and the ``symplectic fermions'' VOA $\mathrm{Fc}$. 

\subsection{C-twist and $\psu(1|1)$}
The C-twist (i.e. the Rozansky-Witten twist) of the theory of a single free hypermultiplet is a 
$\psu(1|1)$ Chern-Simons theory. The simplest holomorphic boundary condition compatible with the C-twist 
supports the $\psu(1|1)$ Kac-Moody VOA, i.e. the symplectic fermions \cite{Kausch} VOA $\Fc$ generated by  fermionic currents with OPE
\be
x(z) y(w) \sim \frac{1}{(z-w)^2}
\ee

The bulk TFT has an algebra of local operators isomorphic to the algebra of polynomials on $\bC^2$, 
i.e. the target of the hypermultiplet theory, with the generators living in cohomological degree $1$. Despite the cohomological degree, it is important to note that these are still \emph{commuting}, not anti-commuting generators.  In this paper we work with objects which have both a $\bZ$-grading  by cohomological degree and a $\bZ/2$ grading by fermion number, and both gradings contribute to signs.  For us it is natural to take the operators of the bulk TFT to be both fermionic and of cohomological degree $1$, and hence commuting.

We will now recover the VOA image of this statement. (See also Section \ref{sec:symplecticfermionext} 
for an equivalent discussion using standard logarithmic VOA naming conventions).

The category $\Fc-\modu$ of finite-dimensional weight modules of $\Fc$ is simple to describe: they are all induced from finite-dimensional modules of the exterior algebra $\Lambda^* \bC^2$ generated by the
current zeromodes $x_0$ and $y_0$. The non-zero modes of the algebra go along for the ride in computations of (derived) morphisms between modules in $\Fc-\modu$, so that they match (derived) morphisms between the corresponding modules of $\Lambda^* \bC^2$.

The self-$\Ext$ algebra of the vacuum module is then computed as the self-$\Ext$ algebra
of the trivial module of $\Lambda^* \bC^2$. This is the same as the Koszul dual of $\Lambda^* \bC^2$,
which is indeed the algebra of polynomials on $\bC^2$ with the generators living in cohomological degree $1$,
precisely as expected!

The two $\Ext_1$ generators are the two extensions of the vacuum module $\cV$ by itself, involving the modules $\cE_\pm$
built from highest weight vectors $|\pm \rangle$ annihilated either by $x_0$ or by $y_0$. To be concrete, the exact sequences take the form 
\begin{equation}
0 \longrightarrow \cV \longrightarrow \cE_\pm \longrightarrow \cV \longrightarrow 0
\end{equation}
where the maps send, say, $|0\rangle \to y_0 |+ \rangle$ and $|+ \rangle \to |0\rangle$. 

We can explicitly compute the full $\Ext$ space by using a projective resolution of the vacuum module. 
The vacuum module $\cV$ has a projective resolution given by the complex $(\bC[u,v] \otimes \cP, x_0 \partial_u + y_0 \partial_v)$, 
where $\cP$ is the projective module generated by a highest weight vector which is not annihilated by either zeromodes.  Its Loewy diagram is given in Figure \ref{fig:LoewFc}. Loewy diagrams are a useful way to visualize the decomposition pattern of modules. We refer to \cite{Creutzig:2013hma} for more details in the example of Fc. 
\begin{figure}[tb]
\begin{center}
\begin{tikzpicture}[scale=0.70][thick,>=latex,
nom/.style={circle,draw=black!20,fill=black!20,inner sep=1pt}
]
\node(left0) at  (1,0) [] {$\cP$:}; 
\node (top1) at (5,2.5) [] {$\cV$};
\node (left1) at (2.5,0) [] {$\cV$};
\node (right1) at (7.5,0) [] {$\cV$};
\node (bot1) at (5,-2.5) [] {$\cV$};
\node(left2) at  (10,0) [] {$\cE_\pm$:}; 
\node (top2) at (12,1.25) [] {$\cV$};
\node (bot2) at (12,-1.25) [] {$\cV$};
\draw [->] (top1) -- (left1);
\draw [->] (top1) -- (right1);
\draw [->] (left1) -- (bot1);
\draw [->] (right1) -- (bot1);
\draw [->] (top2) -- (bot2);
\end{tikzpicture}
\captionbox{\label{fig:LoewFc} The Loewy diagram of the modules $\cE_\pm$ and of the projective cover  $\cP$  of the Fc-vacuum $\cV$.}{\rule{12cm}{0cm}}
\end{center}
\end{figure}

We can write the projective resolution as 
\begin{equation}\label{eq:projsing}
\dots  \longrightarrow \bC^5 \otimes \cP \longrightarrow \bC^4 \otimes \cP\longrightarrow \bC^3 \otimes \cP\longrightarrow \bC^2 \otimes \cP\longrightarrow  \cP \longrightarrow \cV \longrightarrow 0.
\end{equation} 
and apply $\Hom(_,\cV)$. There is a unique map $\cP \to \cV$, which composes to zero with the differentials in the projective resolution, 
giving $\Ext(\cV, \cV) = \bC[u^*,v^*]$, polynomials in two variables of degree $1$.


\subsubsection{C-twist line defects}
This example also offers a good chance to discuss the relation between physical line defects,
topological line defects and VOA modules. 

First of all, we should discuss what do we mean with line defects in the TFT. A 
very broad definition would include all the ways to ``fill in'' a cylindrical hole, i.e. 
all boundary conditions for the circle-compactified bulk theory. A stricter definition 
would only consider local defects in the underlying quantum field theory, whose definition 
only employs a finite number of derivatives of the fields at the location of the line defect.

The two definitions coincide for physical TFTs but not for the sort of cohomological 
TFTs we consider here. The distinction is akin to considering the category of all modules 
for the VOA as opposed to modules which satisfy some finiteness condition on the action 
of sufficiently positive modes of the VOA. 

In this context, the finite-dimensional weight modules of $\Fc$ should be related by line defects 
whose definition only involves the hypermultiplet fields evaluated at the line defect, rather than 
their derivatives. This is a natural choice for line defects which are inherited from renormalizable 
line defects in the original supersymmetric QFT. 

In the physical hypermultiplet theory, maximally supersymmetric line defects compatible with the 
C-twist are built from the pull-back of hyper-holomorphic connections on the target $\bC^2$, aka instantons. 
It should be possible to extend that to more general hyper-holomorphic sheaves. 

For example, a sky-scraper at the origin of $\bC^2$ 
can be described as a complex of trivial bundles: 
\begin{equation}
\cO \to \cO \oplus \cO \to \cO
\end{equation} 
with maps ${z_1 \choose z_2}$ and $(z_2, - z_1)$. A more general instanton bundle has an ADHM construction 
\begin{equation}
\cO^N \to \cO^{2N+M} \to \cO^M
\end{equation} 
with maps 
\begin{equation}
\begin{pmatrix} B_1 + z_1 \cr B_2 + z_2 \cr I \end{pmatrix} \qquad \qquad \qquad \begin{pmatrix} B_2 + z_2 & -B_1 - z_1 & J \end{pmatrix}
\end{equation}
which define a complex iff $[B_1,B_2]+IJ=0$. 

All these examples have the form of complexes $V \otimes \cO$ with differential $d + z_1 d_1 + z_2 d_2$. This gives 
$(V,d)$ the structure of a module for $\Lambda^* \bC^2$, with action given by $d_1$ and $d_2$, which then induces a module for $\Fc$.
According to this identification, the sky-scraper at the origin of $\bC^2$ can be identified with $\cP$. 

In the topologically twisted theory, we can consider more general line defects, associated to sheaves 
on $\bC^2$. Taking $\Ext$ from the sky-scraper sheaf at the origin gives a map to $\Lambda^* \bC^2$ modules, 
as $\Lambda^* \bC^2$ is also the self-$\Ext$ of the sky-scraper sheaf. 
Conversely, a module for $\Lambda^* \bC^2$ gives a complex of sheaves on $\bC^2$ as the cohomology of 
$z_1 d_1 + z_2 d_2$. 

In conclusion, there appears to be a good match between the category of finite-dimensional weight modules of $\Fc$
and a reasonable category of line defects in the bulk TFT. The match takes the form of a Koszul duality.

\subsection{H-twist and symplectic bosons}
The H-twist of the theory of a single free hypermultiplet is expected to control the analytic continuation of a 
symplectic bosons path integral. The symplectic boson VOA $\Sb$ (often also called the $\beta\gamma$-VOA \cite{Ridout:2008nh}) is 
generated by two dimension $\frac12$ bosonic fields $X$ and $Y$ with 
\be
X(z) Y(w) \sim \frac{1}{z-w}
\ee

As the free hypermultiplet has no Coulomb branch, we expect the vacuum module of 
$\Sb$ to also have no self-$\Ext$. Indeed, as the $\Sb$ algebra has no zeromodes
and the vacuum module is just a Verma module, we do not expect non-trivial $\Ext$
algebra. (We refer to section \ref{sec:betagammaext} for a detailed discussion using standard logarithmic VOA naming conventions).

\subsubsection{H-twist line defects}
Maximally supersymmetric defects compatible with H-twist and involving a minimum number of derivatives 
of the hypermultiplet are naturally associated to D-modules on $\bC^2$. Physically, they 
are produced by coupling the hypermultiplets to supersymmetric quantum mechanical systems 
by superpotential couplings $W(X,Y, \cdots)$.  

These D-modules map naturally to modules for $\Sb$ where the first set of negative modes $X_{-\frac12}$ and 
$Y_{-\frac12}$ act as multiplication operators by coordinates on $\bC^2$,
while the first set of positive modes acts by derivatives: we map D-modules on 
$\bC^2$ to modules for two Heisenberg algebras generated by $X_{\pm\frac12}$ and 
$Y_{\pm\frac12}$. This construction can be generalized by including derivatives of the hypermultiplet, 
producing D-modules on the space of negative modes for the symplectic bosons, up to some maximum degree. 

Appropriately defined categories of line defects/D-modules and modules for $\Sb$ will thus coincide, up to the identification between D-modules on $\bC$ and modules for the Heisenberg algebra. 

We must point out that none of these non-trivial line defects/modules
belongs to a class of ``good'' line defects in the physical theory: the required superpotential 
couplings on the line defect are all irrelevant. This is closely related to the fact that the 
module is not highest weight. We will correspondingly exclude them from the category of ``good'' modules, 
which is generated by the vacuum module, with no self-$\Ext$. 

Such line defects can occur as a low energy effective description of an UV line defect in a theory 
which flows to a free hypermultiplet in the IR. In particular, they can play a role in mirror symmetry. 

They may also appear as an ingredient in interacting theories, 
where the hypermultiplet is coupled to gauge fields in such a way that the divergence of gauge invariant 
operators at the line defect is not too severe. The corresponding modules and their extensions 
will play an important role in our gauge theory calculations. 
 
We can give here some particularly interesting, simple example: the infinite-dimensional family of ``spectral flowed modules'' $\sigma_k\left( \Sb\right)$ associated to ``vortex lines'' in the SQFT.
These modules are generated by vectors $|k\rangle$ such that 
\begin{align}
X_{n+\frac12} |k\rangle &=0 \quad n\geq k \cr
Y_{n+\frac12} |k\rangle &=0 \quad n\geq -k
\end{align}
 The $X_{n+\frac12}$, $Y_{-n-\frac12}$ modes in the $\Sb$ VOA form an infinite collection of Heisenberg algebras. 
The vacuum module $\Sb$ and the spectral flow images $\sigma_k\left( \Sb\right)$ 
are the same module for all but a finite collection of such Heisenberg algebras. 
Their extensions can be understood as extensions of Heisenberg modules. 

Given an Heisenberg algebra $\ccH$, generated by $u$,$v$ with $[u,v]=1$, we can define two natural modules $M_u = \bC[[u]]$ and $M_v = \bC[[v]]$ generated by a highest/lowest weight vector annihilated by $u$ or by $v$. There are two natural extensions $\bC((u))$ and $\bC((v))$
which extend the highest weight module by the lowest weight, or viceversa. 

In particular, if we use the Heisenberg algebra generated by $X_{\frac12}$ and $Y_{-\frac12}$
we get an extension from $\Sb$ to $\sigma(\Sb)$ and an extension from $\sigma(\Sb)$ to $\Sb$.

These two natural extensions can be composed into a long exact sequence of modules, starting and ending 
with the vacuum module. This extension is not available in the category of ``good'' modules, but can be discussed in a larger category of modules for the Heisenberg algebra. It gives a self-$\Ext_2$ element of the vacuum module, which we still expect to vanish. It can indeed be shown to be trivial, thanks to the existence of a non-trivial logarithmic module for the Heisenberg algebra which deforms $\Sb \oplus \sigma(\Sb) \oplus \Sb$. Here logarithmic means that the XY current zeromode has non-trivial Jordan blocks. In section \ref{sec:betagammaext} we will discuss these extension of the $\Sb$-algebra and the corresponding categories of logarithmic modules.

\section{The basic Abelian mirror symmetry} \label{sec:hypertwo}
The simplest ${\cal N}=4$ Abelian gauge theory is SQED$_1$, i.e. a $U(1)$ gauge theory coupled to a single 
hypermultiplet of gauge charge $1$. This theory is mirror to the theory of a free hypermultiplet. More precisely, 
SQED$_1$ is a microscopic (aka UV) definition of a theory which flows at large distances (aka in the IR) to 
a theory mirror to a free twisted hypermultiplet. This RG flow should not affect the topologically twisted theory, so 
the C-twist of SQED$_1$ should be equivalent to the H-twist of the free hypermultiplet, and viceversa. 

The simplest holomorphic boundary conditions for SQED$_1$ are also mirror to the simple holomorphic 
boundary conditions for the free hypermultiplet. In particular, the simplest H-twist boundary VOA for 
SQED$_1$ is $\Fc$ and the simplest C-twist boundary VOA for SQED$_1$ is expected to be $\Sb \times \Ff$, 
where $\Ff$ is the holomorphic (spin-)VOA of a complex free fermion (which contributes trivially to the 
category of modules). 

As SQED$_1$ has a free hypermultiplet mirror description and the boundary VOAs are so simple, 
our conjectural relation between bulk lines and modules for SQED$_1$ follows immediately from 
the relation for the free hypermultiplet. On the other hand, looking closely at the mirror dictionary 
can help us identify useful microscopic definitions of gauge theory line defects 
which may be applicable to more general theories and more intricate boundary VOAs.
 
The matching of line defects under mirror symmetry is partly understood \cite{Assel:2015oxa}.  
We will not try to describe here the most general line defects one may define in the gauge theory. 
Notice that there may be non-trivial dynamical identifications between line defects which have different 
microscopic definitions. For example, in the microscopic definition of a standard Chern-Simons theory 
one may define Wilson lines in any representation of the gauge group, 
but the corresponding low energy TFT only has a finite number of simple line defects. 

Another important dynamical subtlety is that a reasonable-looking microscopic definition of a line defect may end up producing 
a defect which is not ``good'' in the sense discussed above. For example, innocent-looking Wilson lines in SQED$_1$ 
will flow to vortex lines in the free hypermultiplet description. A priory, it may also happen that some line defect in the 
IR theory may not have a simple UV definition. 

In the H-twist, SQED$_1$ controls the path integral of a system of gauged symplectic bosons. The boundary VOA is computed 
as a $\mathfrak{u}(1)$-BRST reduction of the product 
\begin{equation}
\Sb \times \Ff
\end{equation}
of a set of symplectic bosons and a set of complex fermions $\psi$ and $\chi$. The BRST charge is built from the diagonal, level $0$ $U(1)$ current in the product VOA. The BRST reduction is generated by BRST-closed bilinears $X \psi \simeq x$ and $Y \chi \simeq y$ and coincides with $\Fc$. A microscopic definition of a line defect can be converted to that of a BRST-invariant module for $\Sb \times \Ff$ and then, passing to BRST cohomology, to a module for the boundary VOA. 

In the C-twist, SQED$_1$ reduces to some Chern-Simons theory based on the $\mathfrak{u}(1|1)$ 
super-algebra at level $1$. The $\Sb \times \Ff$ boundary VOA has a $\mathfrak{u}(1|1)$ current sub-algebra at level $1$ 
defined by bilinears $XY$, $X \psi$, $Y \chi$, $\chi \psi$ of elementary fields. It does not coincide with a $\mathfrak{u}(1|1)$ Kac-Moody VOA, 
but should be thought as an analogue of a WZW model for $\mathfrak{u}(1|1)$. Unfortunately, the map from the 
CS theory description to the boundary VOA involves some non-perturbative calculations which are still poorly understood. 

A given microscopic definition of a line defect can be mapped to the definition of some line in the $\mathfrak{u}(1|1)$ CS theory. 
In turn this will be associated to specific a module for the boundary VOA. This identification may again require non-perturbative calculations
which are currently not understood, or require some amount of guesswork.

\subsection{Bosonization}
Our analysis will be greatly facilitated by a review of an important VOA construction: the {\it bosonization} relationship
between $\Fc$ and $\Sb$.

We can begin by observing the $SL_2$ global symmetry of $\Fc$, which is {\it not} associated to 
any Kac-Moody currents. The $SL_2$-invariant part of $\Fc$ is actually the Virasoro algebra
with $c=-2$ (i.e. $b^2=-2$ in a standard notation), generated by $T = - x y$. Indeed, there is an expansion in $sl_2$ irreps:
\begin{equation}
\Fc = \bigoplus_{n=1}^\infty R^{sl_2}_n \otimes \cV_n^{\Vir}
\end{equation}
where $\cV_n^{\Vir}$ are quotients of Verma modules of dimension $\frac{n(n-1)}{2}$ by the 
submodule generated from the level $n$ null vector.\footnote{This is a degeneration limit of a decomposition 
of the $\mathfrak{osp}(1|2)$ Kac-Moody algebra into modules of the $\mathfrak{su}(2)$ current sub-algebra \cite[]{CREUTZIG2019396,Creutzig:2018ltv}.}

Using the Cartan generator of the $SL_2$ global symmetry, one can define an useful sub-algebra of $\Fc$, 
the zero weight component $\Fc_0$. This is the simplest example $M(2)$ of a ``singlet algebra'' \cite{Kausch:1990vg,Adamovic:2007qs}. 

The other weight components $\Fc_n$ of $\Fc$
give an infinite tower of simple modules of $\Fc_0$. We can write
\begin{equation}
\Fc = \bigoplus_{n=-\infty}^\infty \Fc_n. 
\end{equation}

The $\Sb$ VOA has an $\mathfrak{su}(2)$ current subalgebra at level $-\frac12$ generated by bilinears $X^2$, $XY$, $Y^2$. 
Remarkably, operators in $\Sb$ of weight $0$ under the zeromode of the Cartan current $XY$ can be identify with $\Fc_0 \times \mathfrak{u}(1)$, the product of the singlet VOA and of the VOA generated by the $X Y$ current. 

More generally, we have the decomposition 
\begin{equation}
\Sb = \bigoplus_{n=-\infty}^\infty \Fc_n \otimes \cV^{\mathfrak{u}(1)}_n
\end{equation}
where the $\cV^{\mathfrak{u}(1)}_n \equiv e^{n \varphi}$ are $\mathfrak{u}(1)$ vertex operators of dimension $-\frac{n^2}{2}$, i.e. the $\mathfrak{u}(1)$ current has level $-1$. In other words, $\Sb$ as an infinite simple current extension of $\Fc_0 \times \mathfrak{u}(1)$.

This is just the standard bosonization, familiar from super-string theory textbooks:
\begin{equation}
X = x e^\varphi \qquad \qquad Y = y e^{-\varphi} 
\end{equation}
In other words, $\Sb$ is a sub-algebra of the product of $\Fc$ and a lattice VOA, albeit one with 
unpleasantly negative conformal dimensions. 

\subsubsection{Bosonization and modules}
The bosonization relation maps to a variety of relations between the modules of $\Fc$ and $\Sb$.
These relations may be used to verify or predict relations between the corresponding line defects. 

In the H-twist, a module for the symplectic boson VOA can be combined with the vacuum module of $\Ff$ 
and run through the BRST reduction to get a (possibly dg-) module for $\Fc$. For example, the BRST reduction applied to 
dressed spectral-flowed modules $\sigma_k\left( \Sb\right)\times \Ff$ will result in $\Fc$,
no matter what $k$ is, albeit with a shifted global symmetry grading. 

We can consider extensions of spectral flowed modules, though,
and we will land on extensions of vacuum modules in $\Fc$. For example, the BRST reduction 
applied to the extensions of $\Sb$ by $\sigma(\Sb)$ and of $\sigma(\Sb)$ by $\Sb$ will 
produce $\cE_\pm$!
 
This observation gives us an immediate challenge. The composition of the two extensions 
of $\Sb$ modules produces a long exact extension which could be trivialized in the self-$\Ext_2$ of $\Sb$,
with the help of a certain auxiliary logarithmic module.  
On the other hand, the composition of the two self-$\Ext_1$ generators for $\Fc$ 
is a non-vanishing element in the  self-$\Ext_2$ of $\Fc$. This 
mismatch is likely due to the logarithmic nature of the module mentioned above, which makes it collapse 
upon BRST reduction: the kernel of the current zeromode is smaller then normal in the presence of a 
non-trivial Jordan block. 

Conversely, a module for $\Fc$ can be decomposed by weight and combined with modules of the same charge for $\mathfrak{u}(1)_{-1}$ to induce a module for $\Sb$ and then for $\Sb \times \Ff$. For example,
\begin{equation}
\sigma_k\left( \Sb\right) = \bigoplus_{n=-\infty}^\infty \Fc_n \otimes \cV^{\mathfrak{u}(1)}_{n+k}
\end{equation}
are induced by a degree-shifted image of $\Fc$. Again, extensions of modules in $\Fc$ induce extensions in $\Sb$, but some non-trivial long exact sequences trivialize after the induction. 

These operations, either at the level of gauge theory or at the level of VOA, will have analogues in many of the examples 
we consider through the paper. In general, these methods will allow us to readily produce conjectural generators for the 
$\Ext$ algebra and prove they are indeed generators in some category of modules 
which may be somewhat smaller or bigger than the correct one. We will then have to 
address the more challenging problem to demonstrate that in the correct category of modules
some spurious elements of the $\Ext$ algebra can be set to zero. 

\subsection{Bosonization and line defects}
One basic mirror symmetry expectation is that Wilson lines in SQED$_1$ will map to vortex lines for the 
free hypermultiplet. 

A vortex line of charge $k$, by definition, imposes zeroes and poles on the hypermultiplet which identify it immediately with $\sigma_k(\Sb)$. 
If we dress the highest weight vector $|k \rangle$ by the state of smallest dimension and charge $k$ in the $\Ff$ VOA, 
i.e. $\psi_{-\frac12} \cdots \psi_{\frac12-k}|0\rangle$ for positive $k$ or $\chi_{-\frac12} \cdots \chi_{\frac12+k}|0\rangle$ for negative $k$, 
we obtain a primary field for the $\mathfrak{u}(1|1)$ current algebra with some specific weight proportional to $k$ (see \cite{Creutzig:2008an} for details of this construction). We can tentatively identify 
$\sigma_k(\Sb) \times \Ff$ with the analogue of a WZW primary field associated to a $\mathfrak{u}(1|1)$ Chern-Simons Wilson line of the corresponding 
weight and thus with the corresponding charge $k$ Wilson line in the physical theory, as expected. 

On the other hand, we can seek a definition of line defects in SQED$_1$ which are compatible with an H-twist and 
map, say, to free hypermultiplet Wilson lines associated to some generic sheaf on $\C^2$ or module for $\Lambda^* \bC^2$.
This is a challenging problem in general and we will not attempt to address it here. 

A simpler question is to identify in our language which local operators in SQED$_1$ will match the generators of the self-$\Ext_1$ 
of the vacuum module of $\Fc$. The standard mirror symmetry lore is that the fundamental fields in the free hypermultiplet 
arise as monopole operators in SQED$_1$. We have observed that the $\Ext_1$ generators can be associated to the modules 
$\cE_\pm$, which in turn are produces by the BRST reduction of the basic extensions involving $\Sb$ and $\sigma(\Sb)$. 
The physical interpretation of these extensions is that of microscopic bulk local operators which interpolate between 
the trivial line and a vortex line for the matter fields of SQED$_1$. Up to a singular gauge transformation, this is precisely 
how a gauge theory monopole operator of charge $\pm 1$ looks like! Hence the VOA dictionary is compatible with the standard mirror symmetry dictionary.

\subsection{More on bosonization and $\Ext$} 
There is an analogue of bosonization which applies to boundary VOAs of general theories with Abelian gauge groups.
These include a great majority of the examples we will discuss in the paper. 

Given a theory $T_1$ with a $U(1)$ flavor symmetry acting on the Higgs branch, we can gauge the 
$U(1)$ symmetry to obtain a new theory $T_2$. The theory $T_2$ always has a $U(1)$ flavor symmetry 
acting on the Coulomb branch. Viceversa, gauging that $U(1)$ flavor symmetry of $T_2$ gives back $T_1$. 

This operation can be extended to a relation between certain boundary conditions for $T_1$ and $T_2$ and 
between the corresponding boundary VOAs. In order for our conjecture to hold, it must be the case that these operations on VOAs 
induce a predictable effect on their $\Ext$ algebras, reflecting the relations between the bulk local operators
in $T_1$ and $T_2$.

For example, gauging a $U(1)$ global symmetry in the C-twist of $T_1$ should induce a $U(1)$ symplectic reduction
of the algebra of bulk local operators. At the VOA level, the $U(1)$ gauging operation takes a VOA 
$\cA$ with a global $U(1)_o$ symmetry and produces a new VOA $\cA'$ by dressing operators of weight $n$ with 
appropriate vertex operators of some rank 1 lattice VOA. Can we explain why the self-$\Ext$ of $\cA'$ should be obtained as a $U(1)$ symplectic reduction 
of the self-$\Ext$ of $\cA$?

The $U(1)_o$ symmetry acts on the $\Ext$ algebra of $\cA$. Symplectic reduction projects the $\Ext$ algebra to the $U(1)_o$ weight $0$ sector and quotients away an element of $\Ext_2$ which should play the role of a moment map for the $U(1)_o$.  

Intuitively, the self-$\Ext$ of $\cA_0$ should coincide with the $U(1)_o$ weight $0$ sector of the self-$\Ext$ of $\cA$.
The self-$\Ext$ of $\cA_0 \times \mathfrak{u}(1)$ should include an extra $\Ext_1$ generator, the generator of 
the self-$\Ext$ of $\mathfrak{u}(1)$. The crucial step would be then to prove that the  
operation of extending $\cA_0 \times \mathfrak{u}(1)$ by modules of the form $\cA_n \times \cV^{\mathfrak{u}(1)}_n$ 
will have the effect of turning on a differential which maps the $\Ext_1$ generator for $\mathfrak{u}(1)$ to some ``moment map''
in the $\Ext_2$ of $\cA$, with the net effect of implementing the $U(1)$ symplectic reduction of the $\Ext$ algebra.  
It would be nice to make this expectation precise. 

On the H-twist side, we start from some boundary VOA $\cA'$ with a $\mathfrak{u}(1)$ current algebra at some 
level $-k$. Here $k$ has to be positive in order for the $U(1)$ gauging operation to be possible. 
If it is, we can combine $\cA'$ with some auxiliary holomorphic lattice VOA with a level $k$ $\mathfrak{u}(1)$ current 
and take the $\mathfrak{u}(1)$ BRST reduction to obtain $\cA$. The new Coulomb branch symmetry $U(1)_c$ arises from the 
global charge of the $\mathfrak{u}(1)$ current in the lattice VOA.

The change in the Coulomb branch following a $U(1)$ gauging operation is quite non-trivial. The complex dimension goes up 
by $2$. We should gain a new generator of degree $2$, i.e. a new $\Ext_2$ generator, and add whole new sectors
with non-zero $U(1)_c$ charge to the algebra. 

As the BRST reduction again maps spectral flow modules $\sigma_k(\cA')$ of $\cA'$ to the vacuum module of $\cA$,
we expect the new sectors to arise from the extensions between $\sigma_k(\cA')$ and $\cA'$. 
The origin of the new $\Ext_2$ generator is more obscure. It would be interesting to make this 
discussion more precise. 

\section{A richer example: boundary VOA for $T[SU(2)]$.} \label{sec:tsutwo}
The three-dimensional theory SQED$_2 \equiv T[SU(2)]$ can be defined as the IR limit of a $U(1)$ gauge theory coupled
to two hypermultiplets of charge $1$. It has $SU(2)$ global symmetry acting on the Higgs branch. It also have an $SU(2)$
global symmetry acting on the Coulomb branch, though only the Cartan subgroup is visible int he microscopic description. 
Both Higgs and Coulomb branches are identified with $T^* \bC^2///U(1)$, aka an $A_1$ singularity. 

The theory is conjecturally self-mirror. Furthermore, the same boundary VOA emerges from simple boundary conditions 
compatible with H- and C- twists. 

\subsection{H-twist description of the VOA and modules}
The H-twist description of the boundary VOA is that of a $\mathfrak{u}(1)$-BRST reduction 
of $\Sb^{2} \times \Ff^2$, by the level $0$ $U(1)$ symmetry acting diagonally on all VOAs in the product. 
The bilinears of symplectic bosons and fermions give $\mathfrak{u}(2|2)_1$ currents. 
The BRST reduction removes two Abelian generators, leaving behind
$\mathfrak{psu}(2|2)_1$ currents. 

The VOA is not a $\mathfrak{psu}(2|2)_1$ Kac-Moody VOA, though. For example, the $\mathfrak{su}(2)_1$ current subalgebra 
arises from fermion bilinears and thus it generates the simple quotient $\mathrm{su}(2)_1$ WZW VOA. 
We will denote the boundary VOA as $\mathrm{psu}(2|2)_1$. 

The VOA has an enhanced $SU(2)_o$ global symmetry which rotates the two sets of Grassmann-odd generators 
as a doublet. The whole VOA has a decomposition into modules of $\mathfrak{su}(2)_{-1} \times \mathrm{su}(2)_1 \times SU(2)_o$
of the form \cite{CG17}
\begin{equation}
\mathrm{psu}(2|2)_1 = \sum_{n=0}^\infty \cV_n^{\mathfrak{su(2)}_{-1}} \times \cV_{n \mod 2}^{\mathrm{su}(2)_1} \times V_n^{SU(2)_o}
\end{equation}
where $\cV_n^{\mathfrak{su(2)}_{-1}}$ are Weyl modules associated to zeromode irreps of weight $n$, 
$\cV_{n \mod 2}^{\mathrm{su}(2)_1}$ the irreducible modules for $\mathrm{su}(2)_1$ and $V_n^{SU(2)_o}$
are the $SU(2)_o$ irreps of weight $n$.

Our objective is to compare the self-$\Ext$ of the vacuum module with the algebra of functions on an $A_1$ singularity,
i.e. the Coulomb branch. The $A_1$ singularity has a description as a symplectic quotient $\bC^4//U(1)$. 
The algebra of functions is generated by $U(1)$-invariant bilinears in the coordinate functions on $\bC^4$, modulo the moment map. 
In particular, we expect the vacuum module to have no $\Ext_1$ and have generators in $\Ext_2$ corresponding to these bilinears. 

Notice that the coordinate functions themselves can be seen as sections of a canonical line bundle $\mc{L}$ on $\bC^4//U(1)$
or its inverse. We may hope to find some non-trivial module $M$ for $\mathrm{psu}(2|2)_1$ such that the $\Ext$ group
from $M$ to $\mathrm{psu}(2|2)_1$ coincides precisely with the space of holomorphic sections of $\mc{L}$, and the $\Ext$ group
from $\mathrm{psu}(2|2)_1$ to $M$ coincides precisely with the space of holomorphic sections of $\mc{L}^{-1}$.
Then we could identify the coordinate functions on $\bC^4$ with generators of $\Ext_1(M,\mathrm{psu}(2|2)_1)$ and 
of $\Ext_1(\mathrm{psu}(2|2)_1,M)$ and compose them to identify the desired generators of the self-$\Ext_2$ of the vacuum module. 

We can readily produce interesting modules for $\mathrm{psu}(2|2)_1$ through the BRST construction. 
In particular, consider the BRST reduction of $\Sb \times \sigma(\Sb) \times \Ff^2$. The result is a spectral flowed image of 
the vacuum module of $\mathrm{psu}(2|2)_1$, which we can denote as $\sigma^{(\frac12,\frac12)}(\mathrm{psu}(2|2)_1)$, 
as the flow involves the Cartan generators of both $\mathfrak{su(2)}_{-1}$ and $\mathrm{su}(2)_1$. 

In particular, the $\sigma^{(\frac12,\frac12)}(\mathrm{psu}(2|2)_1)$ module is generated from a vector which is not annihilated by 
the zeromodes of the two bosonic raising generators and an $SU(2)_o$ doublet of fermionic generators. Hence the module 
has a non-trivial action of the zeromodes, but the generator is still annihilated by all positive modes. 
In other words, $\sigma^{(\frac12,\frac12)}(\mathrm{psu}(2|2)_1)$ likely belongs to a category of good modules. 

As spectral flow acts on the basic $\mathrm{su}(2)_1$ modules by exchanging them, 
we have 
\begin{equation}
\sigma^{(\frac12,\frac12)}(\mathrm{psu}(2|2)_1) = \sum_{n=0}^\infty \sigma^{\frac12}\left(\cV_n^{\mathfrak{su(2)}_{-1}}\right) \times \cV_{n+1 \mod 2}^{\mathrm{su}(2)_1} \times V_n^{SU(2)_o}
\end{equation}

We can replace $\sigma(\Sb)$ in the BRST reduction with the extension modules involving $\sigma(\Sb)$ and $\Sb$. 
This should descend to extensions between $\mathrm{psu}(2|2)_1$ and $\sigma^{(\frac12,\frac12)}(\mathrm{psu}(2|2)_1)$.
Furthermore, the same module $\sigma^{(\frac12,\frac12)}(\mathrm{psu}(2|2)_1)$ can also be obtained from BRST reduction of
$\sigma^{-1}(\Sb) \times \Sb$. This gives two distinct extensions between $\mathrm{psu}(2|2)_1$ and $\sigma^{(\frac12,\frac12)}(\mathrm{psu}(2|2)_1)$.

As a result, we find an $SU(2)_o$ doublet $a_{1,2}$ of extensions from $\mathrm{psu}(2|2)_1$ to $\sigma^{(\frac12,\frac12)}(\mathrm{psu}(2|2)_1)$
and a separate $SU(2)_o$ doublet $b^{1,2}$ of extensions from  $\sigma^{(\frac12,\frac12)}(\mathrm{psu}(2|2)_1)$ to $\mathrm{psu}(2|2)_1$.

Assume for a moment that these $\Ext_1$ elements commute in the $\Ext$ algebra and can be combined 
into general self-$\Ext$ for the vacuum module, of the form $a_{i_1} a_{i_2} \cdots b^{j_1} b^{j_2} \cdots$.
This is the algebra of $U(1)$-invariant polynomials in $T^* \bC^2$! In order to get the desired answer, 
we would need to show that the ``moment map'' $\sum_i a_i b^i$ vanishes. 

A very tentative strategy to accomplish that would be to show that although the logarithmic module for 
each $\Sb$ algebra is unsuitable for the BRST reduction, there is some combined $\Sb \times \Sb$ module 
which is logarithmic for each of the $X_i Y^i$ current zeromodes, but is not for the diagonal combination 
which enters the BRST reduction. Such a module could pass through the BRST reduction and 
``certify'' the vanishing in $\Ext_2$ of $\sum_i a_i b^i$.

The physical interpretation of these boundary VOA calculations is transparent. The module $\sigma^{(\frac12,\frac12)}(\mathrm{psu}(2|2)_1)$
corresponds in the gauge theory to a vortex line for the Higgs branch flavor symmetry, defined as a charge $1$ vortex line for one of the two hypermultiplets. 
The generators of $\Ext_1$ are the simplest local operators which can end such a flavor vortex line, i.e. flavor monopole operators. 
These are indeed expected to correspond to the coordinate functions on $\bC^4$. 

There is an alternative possible strategy to show that the $a_i$ and $b^i$ do indeed generate the $\Ext$ algebra. 
The basic idea is to bosonize both symplectic bosons. Then the $\mathrm{psu}(2|2)_1$ is embedded into
a product of $\Fc^2$ and some indefinite lattice VOA, with some bosonization formulae 
recasting each generator into some operator in $\Fc^2$ dressed by appropriate lattice vertex operators. 

We can express this as an infinite simple current extension and do all calculations within the 
category of $\Fc_0^2$ modules. At the end, though, we still need to demonstrate 
the $\sum_i a_i b^i=0$ relation. 

\subsection{The C-twist formulation}
The C-twist of SQED$_2$ gives a CS theory based on a super-algebra with two bosonic Cartan generators and 
four fermionic generators. 

In the gauge theory description of the boundary VOA $\mathrm{psu}(2|2)_1$, a special role is played by 
the sub-algebra generated by the bosonic level $2$ Cartan generator $J$ in $\mathrm{su}(2)_1$ together 
with the level $0$ diagonal combination $I$ of the two Cartan generators in $\mathrm{psu}(2|2)_1$ 
and with the four odd generators which have charge $0$ under $I$: these are the currents which one would predict 
to find at a WZW boundary for that Chern-Simons theory.  The remaining generators arise as boundary monopole 
operators. In this Abelian example, this is a simple current extension. 

The ``flavor vortex line'' which we encountered on the mirror side maps to a charge $1$ Wilson loop 
in the gauge theory, which should also produce $\sigma^{(\frac12,\frac12)}(\mathrm{psu}(2|2)_1)$
at the WZW boundary. 

This makes the tentative construction of the $\Ext_1$ group above even more natural: 
in the C-twist picture, we are trying to reproduce the Higgs branch of the theory
and the bulk local operators which live at the junction between the Wilson loop and the trivial loop 
include the hypermultiplet scalars in homological degree $1$. 

This suggests a general strategy: whenever the bulk algebra of local operators
has an interpretation as the Higgs branch of a gauge theory, the relevant $\Ext$ algebra should be 
generated by the $\Ext_1$ of a collection of modules which correspond to bulk Wilson line defects,
in the same representations as the bulk hypermultiplets. 

\section{Abelian gauge theories} \label{sec:abelian}
Consider an Abelian gauge theory with $N$ hypermultiplets and 
$n$ vectormultiplets. It always has a mirror theory with $N$ hypermultiplets and 
$N-n$ vectormultiplets.

If $Q$ is the $N \times n$ matrix of charges of the hypermultiplets, then 
the mirror theory $N \times (N-n)$ charge matrix $Q^\vee$ is orthogonal to 
$Q$. 

The mirror symmetry is easily understood at the level of the boundary VOA's $\cA$. 
On the H-twist side, we take a $\mathfrak{u}(1)^n-$BRST reduction
of $(\Sb \times \Ff)^N$, using the level $0$ currents
\begin{equation}
\sum_{i=1}^N Q_i^a (\psi_i \chi^i + Y_i X^i)
\end{equation}

Obvious BRST-invariant operators include odd currents 
$M^i = X^i \psi_i$ and $N_i = Y_i \chi^i$ and even currents 
\begin{equation}
J_k = \sum_{i=1}^N (Q^\vee)^i_k \psi_i \chi^i 
\end{equation}
and 
\begin{equation}
\sum_{i=1}^N v_i^k (\psi_i \chi^i + Y_i X^i)
\end{equation} 
with $v_i^k$ defined up to multiples of $Q_i^a$.

These BRST-invariant currents define a VOA $\cA_0$ which is the same as the 
perturbative part of the mirror C-twist boundary VOA. The monopole 
sectors $\cA_q$ on the C-twist side are spectral flow images $\sigma^q \cA_0$
under the $J_k$ currents. 

On the H-twist side these are polynomials in the fermions 
and their derivatives which are mutually local with the currents used in the BRST reduction. 
It is a reasonable conjecture, supported by the equality of H-  and C-twist 
indices/characters, that all other BRST-invariant operators are obtained as the $\cA_0$ image of these. 

We can produce modules $\cW_i$ for $\cA$ by BRST reduction, by replacing the $i$-th 
$\Sb$ vacuum module by $\sigma(\Sb)$. These $\cW_i$ come equipped by construction 
with an $\Ext_1$ to and from the vacuum module. 

On the C-twist side, we identify these pairs of $\Ext_1$ with the $N$ bulk hypermultiplets 
and thus identify $\cW_i$ with Wilson lines of charge $(Q^\vee)^i_k$. Of course, 
$\cW_i$ can also be identified with a spectral flow image of $\cA$. 

Similarly, we can produce modules $\cW_{\vec n}$ where we apply $n_i$ units of spectral flow to the $i$-th
$\Sb$. If $\vec n$ and $\vec n'$ differ only by $1$ at a single location, we have $\Ext_1$ between them, in both directions. 

We conjecture that all self-$\Ext$ of the vacuum module can be produced 
as charge $0$ polynomials in these basic $\Ext_1$'s, modulo 
setting $N-n$ moment maps in $\Ext_2$ to zero. 

\section{A sketch of a proof for all H-twist VOA's} \label{sec:higgs}

We have argued that we can see the algebra of functions on the Coulomb branch of a $3d$ $N=4$ gauge theory by studying the boundary vertex algebra in the $C$-twist.  Braverman-Finkelberg-Nakajima have given a mathematical construction of the Coulomb branch algebra by studying the homology of a certain infinite-dimensional variety, which is closely related to the affine Grassmannian.  In the rest of this section we will give a formal argument that shows how the two constructions are related.  

The discussion below will employ a language which may be less familiar than VOA language for some of the readers. 
The general strategy, though, should be physically transparent. The local operators which parameterize the Coulomb branch 
are monopole operators in the gauge theory. The BFN construction is a mathematical formalization of the construction and multiplication 
of such monopole operators. Abelian examples in the previous sections have taught us the relation between monopole operators and $\Ext$'s built from
the BRST reduction of spectral flowed modules. When the gauge group is non-Abelian, the theory of spectral flowed modules 
is quite rich and poorly investigated in the physics literature. The data which goes into the definition of a spectral flowed module has a close 
relation with the affine Grassmannian. It is natural to expect that a careful calculation of the $\Ext$ algebra employing spectral flowed modules 
will directly reproduce the BFN construction. We will now verify this expectation in some detail.  

Our argument is not quite a mathematical proof.  It relies on a number of non-obvious (to us), but very plausible, statements relating the category of modules for the boundary vertex algebra to categories of $D$-modules on certain infinite dimensional varieties.  

We will assume that we start with a gauge theory with hypermultiplets which live in a holomorphic symplectic representation of the form $R \oplus R^\vee$, where $R$ is some complex representation of the complex gauge group $G$.  We will study the boundary vertex algebra with Neumann boundary conditions for the gauge field and for the hypermultiplets.   As we have seen, we may need to couple to an additional auxiliary holomorphic VOA (usually a collection of free fermions) at the boundary in order to cancel gauge anomalies. For simplicity, we will assume that the auxiliary VOA we introduce at the boundary consists of chiral complex fermions living in a representation $V$ of $G$. Because they are complex fermions, we have two independent boundary fermionic fields, one living in $V$ and one in $V^\vee$.
 
The boundary vertex algebra is a gauged version of the $\beta-\gamma$ system in the vector space $R$, coupled to complex fermions living in $V$.  We will use a flavour symmetry to change the spins of the fields so that the bosonic field $\gamma$ living in $R$ has spin $0$, $\beta$ has spin $1$, the fermion $\psi$ living in $V$ has spin $0$, and the fermion $\psi^\ast$ living in $V^\ast$ has spin $1$.  

If we do this, we can view the boundary algebra as being a gauged $\beta-\gamma$ system living in the super-representation $R \oplus \Pi V$.

Our goal is to relate modules for this vertex algebra to $D$-modules on a certain infinite-dimensional manifold. We will analyze the case where there is no gauge symmetry first.  Suppose we have a non-linear $\beta-\gamma$ system on some complex manifold $X$.  It is known \cite{Witten:2005px, Gorbounov:2016oia} that the algebra of operators of the $\beta-\gamma$ system is the vertex algebra of chiral differential operators on $X$.  

To any vertex algebra $V$ one can assign a (topological) associative algebra $\mc{A}_V$, which is generated by contour integrals of currents.  There is an equivalence of categories between modules for the vertex algebra $V$ and modules for the associative algebra $\mc{A}_V$. This equivalence is strictly true only for the most general possible definition of module. This includes such unphysical modules as those in which the vacuum vector is not annihilated by any of the moes, no matter their spin.  The ``physically reasonable'' modules will be a full subcategory of the category of $\mc{A}_V$-modules.  

Let us view the vertex algebra of chiral differential operators as a sheaf of vertex algebras on $X$. Corresponding sheaf of algebras is the algebra of differential operators on the loop space of $X$. Therefore the category of sheaves of modules for this sheaf of vertex algebras is a full subcategory of the category of $D$-modules on the loop space of $X$. 

We should describe carefully what we mean by the loop space of $X$. We will use the algebro-geometric incarnation of the loop space. This is the space of maps from the formal punctured disc $\what{D}^\times$ to $X$.  Concretely, a map from the formal disc $\what{D}$ to $X$ is an all-order jet of a holomorphic map from $\C$ to $X$.  A map from the formal punctured disc is the same except that we allow finite-order poles at the origin.  For instance, if $X = \C$, then a map from the formal punctured disc to $X$ consist of a Laurent series $\sum a_n t^n$ where $a_n = 0$ for $n \ll 0$. 

In what follows, by $\mc{L} X$ we will always mean the space of maps from the formal punctured disc to $X$.  We will also let $\mc{L}_+ X$ denote those maps with no pole at the origin. 

Under the correspondence between modules for the $\beta-\gamma$ system and $D$-modules on $X$, the vacuum module corresponds to the $D$-module of distributions supported on the subspace $\mc{L}_+ X$. In the terminology of $X$-modules, this $D$-module of distributions is denoted by
\begin{equation} 
\iota_{!} \omega_{\mc{L}_+ X}. 
\end{equation} 
Here $\omega_{\mc{L}_+ X}$ denotes the $D$-module of distributions\footnote{We are using right $D$-modules instead of left $D$-modules.  In finite dimensions there is no essential difference, but in infinite dimensions there is. A left $D$-module can be viewed as a differential equation that can be satisfied by a function.  Just like functions, left $D$-modules work well under pull-back. Right $D$-modules are differential equations that can be satisfied by distributions (including smooth top forms). Like distributions, right $D$-modules behave well under push-forward.} on $\mc{L}_+ X$, and $\iota_!$ is a push-forward operation which takes a $D$ module on $\mc{L}_+ X$ to one on $\mc{L} X$.

\subsection{Gauged $\beta-\gamma$ systems and CDOs on stacks}
Suppose we have a group $G$ acting on $X$, and we consider the gauged $\beta-\gamma$ system on $X$.  This gauged $\beta-\gamma$ system on $X$ behaves exactly like the $\beta-\gamma$ system on the quotient stack $X/ G$.  The most important point to understand is that the algebra of currents for the gauged $\beta-\gamma$ system behaves like differential operators on the loop space of $X / G$.  To understand differential operators on this space, we should first understand its classical limit, which is functions on the cotangent bundle of the loop space of $X / G$.

The cotangent bundle of the loop space of $X / G$ is the same as the loop space of the cotangent bundle. Further, the cotangent bundle of the stack $X / G$ is the symplectic reduction of $T^\ast X$ by $G$. It is essential, however, that we use the \emph{derived} symplectic reduction.  In the derived symplectic reduction, instead of simply setting the moment map to zero, we introduce a fermionic variable $\b$ living in the adjoint representation of $G$ with a differential 
\begin{equation} 
\d \b_a = \mu_a 
\end{equation}  
(where $\mu$ is the moment map).

We therefore find that the cotangent bundle to $X / G$ is the quotient of the dg manifold $\g^\ast [-1] \oplus T^\ast X$ by $G$, where the coordinate on $\g^\ast[-1]$ is denoted $\b$. 

The loop space of this is the quotient of the loop space of $\g^\ast[-1] \oplus T^\ast X$ by the loop space of $G$. If we introduce local coordinates $\gamma,\beta$ on $T^\ast X$, where $\gamma$ is a coordinate on the base and $\beta$ on the fibre, we find that the algebra of functions on the loop space of $\g^\ast [-1] \oplus T^\ast X$ can be described in terms of polynomials of $\b_n,\gamma_n, \beta_n$ which are the modes associated to the coordinates $\b,\gamma,\beta$.  These operators have a BRST differential
\begin{equation} 
\d \b_{a,n} =  \mu(\beta,\gamma)_{a,n} 
\end{equation} 
where the mode $\b_{a,n}$ of $\b_a$ is sent to the corresponding mode of the moment map.  
 
We should further restrict to those operators which are invariant under the action of the loop group.  Since the loop group is not semi-simple, we should take the derived functor of invariants, instead of the naive invariant operators. This is achieved by taking the Lie algebra cochains of the loop algebra $g((z))$ with coefficients in the dg module given by the polynomials of the $\b_n,\beta_n,\gamma_n$.   Forming Lie algebra cochains amounts to introducing a second sequence of fermionic variables $\c_{a,n}$ which transform in the co-adjoint representation of $\g$.  These, of course, are the familiar $\c$-ghosts and are involved in the BRST operator in the usual way:
\begin{align} 
\d \b_{a,n} &= f_{abc} \sum_{r+s=n} \c_{b,r}\b_{c,s}\\
\d \beta_{i,n} &= \sum_{r+s = n} \c_{a,r} \frac{\partial}{\partial \gamma^i_{-n}}  \mu(\beta,\gamma)_{a,r} \\
 \d \gamma^i_{n} &= \sum_{r+s = n} \c_{a,r} \frac{\partial}{\partial \beta_{i,-n} } \mu(\beta,\gamma)_{a,r} \\ 
\d \c_{a,n} &= f_{abc} \sum_{r+s = n} \c_{b,r} \c_{c,s}.
\end{align} 
Further, since the ghosts $\c_{a}$ are associated to the tangent space of the stack $B G$, and the ghosts $\b_a$ to its cotangent space, we find they are canonically conjugate, as expected.

This argument shows that the phase space for the $\beta-\gamma$ system on the symplectic reduction stack $T^\ast(X)//G$ coincides with that obtained by adjoining $\b$ and $\c$ ghosts to the $\beta-\gamma$ system.  It follows that the algebra of currents for the gauged $\beta-\gamma$ system coincides with a quantization of the algebra of functions on the loop space of $T^\ast(X) // G$. Since this quantization behaves well with respect to scaling of the cotangent fibres, it deserves to be called the algebra of differential operators on the loop space of $X / G$.

\subsection{The affine Grassmannian and endomorphisms of the vacuum module}
This argument implies that modules for the gauged $\beta-\gamma$ system are $D$-modules on the loop space of the stack $X / G$. These are the same as (strongly) $\mc{L} G$ equivariant $D$-modules on $\mc{L} X$. 

We are interested the case when $X = R \oplus \Pi V$ is a super-representation of $G$.  Our aim is to use this description of the category of modules for the gauged $\beta-\gamma$ system to argue that the self-Ext's of the vacuum module is the algebra of functions on the Coulomb branch as constructed by \cite{Braverman:2016wma}. The first step in the argument is to show that we can dispense with the fermionic representation $\Pi V$, which we introduced to cancel the anomaly in the vertex algebra.  In finite dimensions, if $\Pi W \to Y$ is a fermionic vector bundle on a manifold $Y$, then the category of $D$-modules on the total space of $\Pi W$ is equivalent to the category of $D$-modules on $Y$. Indeed, the algebra $D(\Pi W)$ is locally the tensor product of differential operators on $Y$ with the Clifford algebra of $W \oplus W^\vee$. This Clifford algebra is Morita trivial, leading to the equivalence of categories.  Concretely, the equivalence of categories is realized by the push forward along the zero section $Y \into \Pi W$.  Locally, this push-forward sends a $D$-module $M$ on $Y$ to $M \otimes \wedge^\ast W$, the tensor product of $M$ with the irreducible representation of the Clifford algebra on $W \oplus W^\vee$. 

It is reasonable to posit that this holds true in infinite dimensions as well.   If so, we would expect that the push-forward along the inclusion map
\begin{equation} 
\mc{L} ( R / G) \to \mc{L} (R \oplus \Pi V)/G  
\end{equation}
gives rise to an equivalence of categories of $D$-modules.

Justification for this is provided by the fact that the category of representations of the complex free fermion vertex algebra is equivalent to the category of vector spaces.  This is the analog of the statement that the Clifford algebra is Morita trivial.

One can ask why it is reasonable to consider $D$-modules on the loop space of $R / G$ even if the gauged $\beta-\gamma$ system is anomalous.  The anomaly in the gauged $\beta-\gamma$ system will imply that the algebra of differential operators on $R / G$ will be ill-defined: the BRST operator that will appear in its definition will not square to zero.  This does not, however, imply that the \emph{category} of $D$-modules on $\mc{L} (R / G)$ is ill-defined. It only tells us that the $D$-module $D_{\mc{L}(R / G)}$ is not defined.   

\subsection{Computing self-Ext's in $D$-module language}
Given this, let us compute the self-Ext's in the category of $D$-modules on $\mc{L}(R / G)$ of the $D$-module
\begin{equation} 
\iota_! \omega_{\mc{L}_+(R / G)} 
\end{equation}
where $\mc{L}_+(R / G)$ is the space of maps from the formal disc to $R / G$.  To do this computation, we will use a result established by Chriss and Ginzburg \cite{CG} in finite dimensions, which we posit also holds in infinite dimensions.  

Chriss and Ginzburg show the following. Suppose that $f : Z \to Y$ is a map and we consider the $D$-module $f_! \omega_Z$ on $Y$.  Then, the self-Ext's of $f_! \omega_Z$ can be computed as
\begin{equation} 
\op{Ext}^\ast_{D(Y)-\op{mod}}(f_! \omega_Z, f_! \omega_Z) = H_\ast( Z \times_Y Z). 
\end{equation}
On the right hand side we have the homology of the fibre product of $Z$ with itself over $Y$, which is an associative algebra under convolution. This isomorphism is an isomorphism of algebras.

Applying this to our infinite-dimensional situation, we conclude that
\begin{equation} 
 \op{Ext}^\ast_{D(\mc{L}(R/G))} ( \iota_! \omega_{\mc{L}_+(R / G)}, \iota_! \omega_{\mc{L}_+(R / G)} ) = H_\ast ( \mc{L}_+ (R / G) \times_{\mc{L}(R / G)} \mc{L}_+ (R / G) ). 
\end{equation}

\subsection{Connecting with the work of Braverman, Finkelberg and Nakajima}
Finally we need to relate the space $ \mc{L}_+ (R / G) \times_{\mc{L}(R / G)} \mc{L}_+ (R / G) )$ to the moduli spaces studied by Braverman, Finkelberg and Nakajima.   A point in the space $  \mc{L}_+ (R / G) \times_{\mc{L}(R / G)} \mc{L}_+ (R / G) )$ consists of two maps $\phi_1, \phi_2 : D \to R / G$, with a gauge transformation relating them on the punctured formal disc $D^\times$.  We can describe this data in more detail. It consists of: 
\begin{enumerate} 
\item $r_1 \in R[[z]]$, a representative for the map $\phi_1 : D \to R / G$.  
\item $r_2 \in R[[z]]$, which is a representative for the map $\phi_2 : D \to R / G$. 
\item A gauge transformation $g \in G((z))$ such that\footnote{Strictly speaking, this equation should be imposed at the derived level -- meaning that odd variables should be introduced whose differential imposes the relation. We will not be concerned about the difference between the derived and underived versions  of the space: for one thing, these differences do not affect homology.}  $g \cdot r_1 = r_2$. 
\end{enumerate}
This data is taken up to the action of a pair of elements $\rho_1,\rho_2 \in G[[z]]$, which act by
\begin{align} 
r_1  &\mapsto \rho_1 \cdot r_1 \\
r_2 & \mapsto \rho_2 \cdot r_2 \\
g & \mapsto \rho_2\cdot  g\cdot \rho_1^{-1}.  
\end{align} 
If we only take the quotient by one copy of $G[[z]]$, say that given by $\rho_1$, we find the space of triples introduced by Braverman, Finkelberg and Nakajima \cite{Braverman:2016wma}. The quotient by the action of the second copy of $G[[z]]$ produces a stack, whose homology can be modelled by the $G[[z]]$-equivariant homology of the space of triples.

We conclude that the self-Ext's of the $D$-module $\iota_! \mc{L}_+ (R / G)$ should be the $G[[z]]$-equivariant homology of the space of triples, precisely as in \cite{Braverman:2016wma}. The general results of Chriss and Ginzburg \cite{CG} tell us that the product on the Ext groups are computed in terms of convolution, which in this case is the convolution product used in \cite{Braverman:2016wma}.  We conclude that there should be an isomorphism of algebras between the self-Ext's of the vacuum module for our vertex algebra and the algebra defined in \cite{Braverman:2016wma}.  
\subsection{The problem of $D$-affineness}
How close is this argument to being a proof?  We are very far from being experts on the theory of $D$-modules on infinite-dimensional varieties. Even so, it seems plausible that many of the arguments we have borrowed from the theory of $D$-modules on finite-dimensional varieties will probably work, when appropriate care is taken, in the infinite-dimensional case too.  

There is one issue, that seems to us more serious, and restricts the domain of validity of our conjecture.  Given a variety (or stack) $X$, we said that we should identify the category of modules for the vertex algebra of chiral differential operators on $X$ with the category of $D$-modules on the loop space of $X$.  This is true as long as we treat chiral differential operators as a \emph{sheaf} on $X$, and consider sheaves of modules. It is not at all obvious that the category of modules for the global sections of the sheaf of vertex algebras on $X$ will be equivalent to the category of $D$-modules on the loop space of $X$.  

With ordinary differential operators instead of chiral differential operators, a variety $X$ is called \emph{$D$-affine} if the category of sheaves of modules over the sheaf $D_X$ of algebras on $X$ is equivalent to the category of modules over the global sections algebra $\Gamma(X, D_X)$. Being $D$-affine is strictly weaker than being affine.  

Similarly, in the chiral world, we say that $X$ is \emph{chiral $D$-affine} if the category of sheaves of modules over the sheaf of vertex algebras $D_X^{ch}$ is equivalent to the category of modules over its global sections vertex algebra $\Gamma(X, D_X^{ch})$.   

Since the vertex algebras we consider in the body of the paper are global objects, and not sheaves on the Higgs branch, the argument relating these vertex algebras to the construction of \cite{Braverman:2016wma} can only work if the Higgs branch (supplemented by fermions to cancel the anomaly) is chiral $D$-affine.  

We will phrase a conjecture concerning when we expect this to hold.
\begin{conjecture}
Suppose that $R$ is a representation of a semi-simple group $G$ such that for $t$ in the Lie algebra of the compact form of $G$, we have 
\begin{equation} 
\op{Tr}_R (t^2) > \op{Tr}_{\g} (t^2). 
\end{equation}
Let $V$ be any representation of $G$ such that 
\begin{equation} 
 \op{Tr}_R (t^2) - \op{Tr}_{\g} (t^2) =   \op{Tr}_V (t^2) 
\end{equation}
Then the stack quotient $(R \oplus \Pi V)/G$ is chiral $D$-affine. 
\end{conjecture}

\section{A general argument for C-twist boundary VOA's in perturbation theory} \label{sec:coulomb}
The C-twist VOA for Dirichlet boundary conditions should be defined non-perturbatively 
as some kind of WZW model for a Chern-Simons theory based on the super-algebra introduced in 
\cite{Costello:2018fnz}. Mathematically, the WZW construction should present the VOA as 
the homology of some bundle on the affine Grassmanian.

In perturbation theory, one only obtains the Kac-Moody current algebra for that super-algebra. The 
self-$\Ext$ of the perturbative C-twist VOA is actually already surprisingly close to the desired answer,
i.e. the algebra of functions on the Higgs branch. 

As we verify in appendix \ref{app}, only the zero modes of Kac-Moody superalgebra contribute to  the self-$\Ext$'s. These yield a Chevalley complex with the following generators:
\begin{enumerate} 
	\item	A ghost number $1$ odd generators $c \in \mathfrak{g}^*$, $b \in \mathfrak{g}$ 
	\item Ghost number $1$ even generators $q$ in the hypermultiplet representation.  
\end{enumerate}
The differential is 
\begin{align}
\delta c &= [c,c] \cr
\delta q &= [c,q] \cr
\delta b &= \mu(q) + [c,b]
\end{align} 
where $\mu(q)$ is the moment map. 

The cohomology of this complex can be computed in two steps: first, we can ignore $c$ and take the cohomology only of the term in the differential $\delta b = \mu(q)$.   This complex is the Koszul complex for the equations $\mu_a(q) = 0$ (where $a$ is a Lie algebra index).  The cohomology of this complex includes polynomials in $q$, modulo those which vanish on the sub-variety $\mu(q) = 0$. 

In principle, there may also be cohomology classes involving $b$.  However, under the hypothesis that $\mu(q) = 0$ is a codimension $\op{dim} \g$ subvariety of the hypermultiplet representation, there are no such cohomology classes.  This implies that coefficients  $\mu_a(q)$ of the moment map form a ``regular sequence'', so that there is no higher cohomology in the Koszul complex. 

Therefore, under this mild hypothesis, taking cohomology of this term in the differential produces the algebra of functions on the zero-locus of the moment map.

To find the Higgs branch, we also want to restrict to gauge invariant polynomials.  Taking the cohomology of the terms in the differential involving $c$ does this, but also introduces some extra unphysical operators.  The unphysical operators are those involving just $c$, such as $f_{a_1a_2 a_3} c_{a_1} c_{a_2} c_{a_3}$.  These extra operators form a copy of the Lie algebra cohomology of $\g$, which is isomorphic to the cohomology of the Lie group $G$.  

In the end, we find that the self-$\Ext$'s of the Kac-Moody superalgebra produce 
\begin{equation} 
	\mathcal{O}(\mc{M}_H) \otimes H^\ast(G), 
\end{equation}
the tensor product of functions on the Higgs branch with the cohomology of the group $G$.

We hope that a non-perturbative analysis, involving boundary monopoles, will cancel the factor of $H^\ast(G)$.

\section{Non-Lagrangian generalizations} \label{sec:ad}
The VOA $\Fc$ is the first member of a remarkable family of VOAs, 
the triplet VOAs $W(p)$.\footnote{Strictly speaking, the triplet VOA $W(2)$ is the $Z_2$-even part of 
$\Fc$. In general, even triplet VOAs $W(2n)$ admit extensions analogue to $\Fc$.}

Four-dimensional gauge theory constructions along the lines of \cite{CG17} give some good reasons to 
believe that triplet VOAs are indeed C-twist boundary VOAs for some rather mysterious 
3d ${\mathcal N}=4$ SCFTs, defined implicitly as the boundary degrees of freedom 
for some boundary conditions of 4d $SU(2)$ ${\mathcal N}=4$ SYM, which in turns are defined 
as the images of Neumann boundary conditions under certain duality transformations \cite{CG17}. 

In the language of \cite{Gaiotto:2008sd, Gaiotto:2008ak} the 3d theories can be thought off as resulting from 
gauging the Coulomb branch $SU(2)$ symmetry of $T[SU(2)]$, with $p$ units of Chern-Simons coupling. 
The Higgs branch should be unaffected by this and remain an $A_1$ singularity.  

We would like to identify the C-twist VOA with the $W(p+1)$-triplet VOA. This is the VOA for the best understood logarithmic CFTs. 
We use results of Adamovi\'c and Milas \cite{AM1, AM2} (another important early work is \cite{Feiginetal}). 

The simple modules are denoted by $W_{s, r}$ with integer labels $1\leq s\leq p$ and $1\leq r\leq 2$. 
The $W_{p, r}$ are projective and hence don't have any extensions. 
\begin{figure}[tb]
\begin{center}
\begin{tikzpicture}[scale=0.70][thick,>=latex,
nom/.style={circle,draw=black!20,fill=black!20,inner sep=1pt}
]
\node(left0) at  (0,0) [] {$R_{s,r}$:}; 
\node (top1) at (5,2.5) [] {$W_{s, r}$};
\node (left1) at (2.5,0) [] {$W_{p-s, 3-r}$};
\node (right1) at (7.5,0) [] {$W_{p-s, 3-r}$};
\node (bot1) at (5,-2.5) [] {$W_{s, r}$};
\draw [->] (top1) -- (left1);
\draw [->] (top1) -- (right1);
\draw [->] (right1) -- (bot1);
\draw [->] (left1) -- (bot1);
\end{tikzpicture}
\captionbox{\label{fig:loewytriplet} Loewy diagram of the projective cover $R_{s, r}$ of the simple triplet module $W_{s, r}$ for $1\leq s\leq p-1$.}{\rule{12cm}{0cm}}
\end{center}
\end{figure}
Projective covers of the others have Loewy diagram as in Figure \ref{fig:loewytriplet}. Let $\omega$ be the fundamental weight of $\mathfrak{sl}_2$ and $\rho_{n\omega}$ the irreducible highest-weight representation of highest-weight $n\omega$. 
We find in section \ref{sec:triplet} the projective resolution
\begin{equation}\label{eq:projtrip2}
\dots \longrightarrow 4 R_{p-s, 3-r} \longrightarrow 3 \otimes R_{s, r}\longrightarrow 2 R_{p-s, 3-r} \longrightarrow  R_{s, r} \longrightarrow W_{s, r} \longrightarrow 0.
\end{equation} 
which suggests an $SU(2)$ action, i.e. 
\begin{equation}\label{eq:projtrip2}
\dots \longrightarrow \rho_{3\omega} \otimes R_{p-s, 3-r} \longrightarrow \rho_{2\omega} \otimes R_{s, r}\longrightarrow \rho_{\omega} \otimes R_{p-s, 3-r} \longrightarrow \rho_{0} \otimes R_{s, r} \longrightarrow W_{s, r} \longrightarrow 0.
\end{equation}
Let $\mathbb Z_2:=\mathbb Z/2\mathbb Z$ and consider its action on $\mathbb C[x, y]$ induced by mapping $x, y$ to $-x, -y$ so that $\mathbb C[x, y]$ decomposes in eigenspaces as
$\mathbb C[x, y] = \mathbb C[x, y]^{\mathbb Z_2} \oplus \mathbb C[x, y]_-$.
We see that 
\begin{equation}
\text{Ext}^\bullet_{\mathcal C}(W_{s, r}, W_{s, r}) \cong \mathbb C[x, y]^{\mathbb Z_2}, \qquad\qquad \text{Ext}^\bullet_{\mathcal C}(W_{s, r}, W_{p-s, 3-r}) \cong \mathbb C[x, y]_-
\end{equation}
and so the spectrum of this Ext-ring is the $A_1$-singularity
\begin{equation}
\text{Spec}\left( \text{Ext}^\bullet_{\mathcal C}(W_{s, r}, W_{s, r})\right)  = \mathbb C^2/\mathbb Z_2.
\end{equation} 

We may also attempt to propose an H-twist VOA for these 3d theories. Work in progress on 4d $SU(2)$ ${\mathcal N}=4$ SYM suggests that the 
H-twist VOA should be $L_{p-3/2}(\mathfrak{sl}_2)$. Based on examples, the ``associated variety'' of an H-twist boundary VOA is expected to coincide with the Higgs branch of the 3d theory. It turns out that the $A_1$-singularity is the associated variety of the affine VOA of $\mathfrak{sl}_2$ at any admissible level \cite{arakawa2016}. 

The associated variety of the triplet $W(p)$ is just a point as it is a $C_2$-cofinite VOA. This suggests that the Coulomb branch of these 3d theories
should simply be a point. Given that the representation theories of all admissible level $L_{k}(\mathfrak{sl}_2)$ \cite{AdaVer95,Creutzig:2013yca} are quite similar to the one of $L_{-1/2}(\mathfrak{sl}_2)$  \cite{Ridout:2008nh, Creutzig:2012sd}, 
it is plausible that the Ext-algebra of $L_{p-3/2}(\mathfrak{sl}_2)$ may also be trivial (in the category of all finite length modules).

These conjectures have natural generalizations for other gauge groups as well and the natural higher rank generalization to the triplet VOAs are the logarithmic W-algebras of Boris Feigin and Tipunin \cite{Feigin:2010xv,Lentner:2017dkg}. The higher rank analogues of the singlet algebra are called narrow W-algebras \cite{Creutzig:2016uqu} and extensions of these narrow W-algebras times Heisenberg VOAs appear in the context of higher rank Argyres-Douglas theories \cite{Creutzig:2018lbc}.

\section{Controlling $\Ext$ algebras by simple current extensions} \label{sec:voa}

We will now compute $\Ext$ algebras for various VOAs. All these VOAs have in common that they can be realized as simple current extensions of singlet algebras and Heisenberg algebras. Firstly, we state the result of the computations. The relevant VOAs are then introduced below.  

Let $M(p)$ be the singlet VOA, $W(p)$ the triplet and $L_1(\mathfrak{psl}(N|N))$ the simple affine super VOA of $\mathfrak{psl}(N|N)$ at level one and $\widetilde {L_1(\mathfrak{psl}(N|N))}$ a certain simple current extension of $L_1(\mathfrak{psl}(N|N))$ which appears as H-twist VOA for SQED$_N$. Remark that in the cases $N=1, 2$ $\widetilde {L_1(\mathfrak{psl}(N|N))}$ is just ${L_1(\mathfrak{psl}(N|N))}$ while for $N>2$ it is an infinite order extension of ${L_1(\mathfrak{psl}(N|N))}$.

For each VOA we will have two categories $\mathcal C\subset \mathcal C_{\text{log}}$. By $\mathcal C$ we mean the category whose objects are subquotients of iterated tensor products of simple objects and by $\mathcal C_{\text{log}}$ we mean the category of all finite length objects. The results depend very much on the chosen category and we will see that in the larger category $\mathcal C_{\text{log}}$ we have more relations, i.e. equivalences of chain complexes, between extensions and so we get smaller Ext-algebras. It turns out that these latter smaller ones compare nicely to our expectations from gauge theory considerations. 

\subsection{The Result}

Our result is as follows: 

Assuming correctness of Conjecture 5.1 and Conjecture 5.8 of \cite{CGR} we have the following $\text{Ext}^\bullet(V, V)$-algebras. Firstly in $\cat$
\begin{enumerate}
\item $\text{Ext}^{\bullet}_{\cat}\left(M(p), M(p)\right) \cong \mathbb C[x^2]$; 
\item $\text{Ext}^{\bullet}_{\cat}\left(W(p), W(p)\right) \cong \mathbb C[x^2, y^2, xy]$; 
\item $\text{Ext}^{\bullet}_{\cat}\left(\widetilde{L_1(\mathfrak{psl}(N|N))}, \widetilde{L_1(\mathfrak{psl}(N|N))}\right) \cong \mathbb C[x_1^2, \dots, x_N^2, v_\pm^N]/f$; 
\end{enumerate}
with $f=x_1^2x_2^2\dots x_N^2-v_-^Nv_+^N$
and secondly in $\catlog$
\begin{enumerate}
\item $\text{Ext}^{\bullet}_{\catlog}\left(M(p), M(p)\right) \cong \mathbb C$; 
\item $\text{Ext}^{\bullet}_{\catlog}\left(W(p), W(p)\right) \cong \mathbb C[x^2, y^2, xy]$; 
\item $\text{Ext}^{\bullet}_{\catlog}\left(\widetilde{L_1(\mathfrak{psl}(N|N))}, \widetilde{L_1(\mathfrak{psl}(N|N))}\right) \cong \mathbb C[x^2, v_\pm^N]/f$ 
\end{enumerate}
with $f=x^{2N}-v_-^Nv_+^N$. This is the algebra of functions on the Coulomb branch of SQED$_N$, as expected.

\subsection{The Argument}

Computing the Ext-algebras splits into several steps. The triplet VOA is an extension of the singlet VOA and $L_1(\mathfrak{psl}(N|N))$ is an extension of $M(2)^{\otimes N}\otimes L_1(\mathfrak{sl}_N)\otimes H^{\otimes (N-1)}$ with $H^{\otimes m}$ the Heisenberg VOA of rank $m$. In categorical terms, this means that the category of modules for the extended VOA is the category of local modules for the corresponding superalgebra object in the category of modules of the smaller VOA. One can then study the category of modules of the small VOA and use induction to obtain desired results in the big VOAs. We thus need to understand the singlet VOAs. Unfortunately not everything is known there and so the argument will rely on the fairly well studied conjecture of braided equivalence of weight modules of unrolled restricted quantum groups and singlet VOAs. 

We will now follow this outline step by step. 
\begin{enumerate}
\item In section \ref{sec:VOAextensions} we will discuss VOA-extensions of simple current type and explain in two prototypical Examples how projective resolutions and thus $\Ext$-algebras behave under VOA-extension. 
\item Our most important basic building block VOA is the singlet algbera $M(p)$ for $p\in \mathbb Z_{>1}$ and especially for $p=2$. In section \ref{sec:singlet} we thus compute $\Ext$ algebras for simple $M(p)$-modules. This is the main computational step. 
\item There is a straightforward lift of these results to $\Ext$ algebras of modules for multiple copies of $M(p)$, presented in \ref{sec:manysinglets}. 
In particular, we apply the example of section \ref{ex:two} to a diagonal simple current extension of many copies of $M(p)$.
\item The example of section \ref{ex:two} can be applied to the triplet VOA $W(p)$ as a simple current extension of $M(p)$ and thus in section \ref{sec:triplet} we get the $\Ext$ algebras of simple triplet modules as corollary. 
\item In section \ref{sec:pslNN} we discuss $L_1(\mathfrak{psl}(N|N))$ and its simple current extension $\widetilde{L_1(\mathfrak{psl}(N|N))}$. Both can be realized as VOA-extensions of $N$-copies of $M(2)$ and the example of section \ref{ex:one} immediately applies. 
\end{enumerate}

\subsection{Vertex algebra extensions}\label{sec:VOAextensions}

Let $V$ be a VOA and let $\mathcal C$ a full vertex tensor category of $V$-modules in the sense of \cite{HLZ1, HLZ2, HLZ3, HLZ4, HLZ5, HLZ6, HLZ7, HLZ8}. Let $V\subset W$ be a bigger VOA containing $V$ exactly once, such that $W$ is an object of $\mathcal C$. Then $W$ defines a commutative, associative, haploid algebra $A$ in $\mathcal C$ \cite[Thm. 3.2. and Rmk. 3.3]{HKL} and the category $\catA$ of local $A$-modules in $\mathcal C$ is braided equivalent to the category of $W$-modules in $\mathcal C$ \cite[Thm. 3.65]{CKM}. Moreover there is a functor $\mathcal F$ from $\mathcal C$ to $A$-modules. The induced object as a $\mathcal C$-module is
\[
\mathcal F(X) \cong_{\mathcal C} A\boxtimes_{\mathcal C} X.
\]
An object $X$ of $\mathcal C$ is called a simple current if it is invertible in the tensor ring. $W$ is called a simple current extension of $V$ if it is a direct sum of inequivalent simple currents of $V$. Assume that $\mathcal C$ is rigid and that $W$ is a simple current extension. Assume that $W$ is simple as a module for itself. The induction functor $\mathcal F$ is exact (tensor product is bi-exact because of rigidity) and maps simple to simple modules \cite[Prop. 4.5]{CKM} and projective to projective modules \cite[Rmk. 2.64]{CKM}. Especially it preserves projective resolutions, i.e. let $M$ be an object in $\mathcal C$ and let 
\begin{equation}\nonumber
\dots  \longrightarrow P^4_M\longrightarrow P^3_M\longrightarrow P^2_M\longrightarrow P^1_M\longrightarrow P^0_M \longrightarrow M \longrightarrow 0
\end{equation} 
be a projective resolution of $M$, then 
\begin{equation}\nonumber
\dots  \longrightarrow \mathcal F(P^4_M)\longrightarrow  \mathcal F(P^3_M)\longrightarrow  \mathcal F(P^2_M)\longrightarrow  \mathcal F(P^1_M)\longrightarrow  \mathcal F(P^1_M) \longrightarrow  \mathcal F(M) \longrightarrow 0
\end{equation}
is a projective resolution of $\mathcal F(M)$ in the category $\mathcal C_A$ of $A$-modules in $\mathcal C$. Now assume that $\mathcal F(M)$ is in the category $\catA$ of local $A$-modules and also assume that every object of $\mathcal C$ is a subquotient of iterated tensor products of simples in $\mathcal C$ then by \cite[Thm. 3.20]{CKL} all the projectives $\mathcal F(P^n_M)$ are local as well, i.e. we have obtained a projective resolution of the $W$-module $\mathcal F(M)$ in the category of $W$-modules that lie in $\mathcal C$. 
Let us illustrate this in the two types of situations that we need:

\subsubsection{Example 1}\label{ex:one}

Simple current extensions beyond semi-simplicity are studied in \cite{CKL} and we refer to that work for further details. 
Let $V_1$ and $V_2$ be two VOAs with rigid vertex tensor categories $\mathcal C_1$ and $\mathcal C_2$. Let $\mathcal C=\mathcal C_1\boxtimes \mathcal C_2$. We assume that $\mathcal C_1$ is semi-simple and every simple object of $\mathcal C_1$ is invertible, i.e. a simple current. In other words there is an abelian group $G=(G, \cdot, e)$, such that simple objects $J_g$ of  $\mathcal C_1$ are labelled by elements $g$ of $G$ and tensor product is
\[
J_g \boxtimes_{\mathcal C_1} J_{h} \cong J_{g\cdot h}.
\]
The VOA $V_1$ is the tensor identity $J_e$. 
Heisenberg and lattice VOAs are prototypical examples for $V_1$. 
Let $H\subset G$ be a subgroup of $G$ and let $\mathcal C_2$ contain a full tensor subcategory $\mathcal C_H$ whose simple objects are invertible and inequivalent simples $M_g$ are labelled by elements $g$ of $H$, s.t.   $M_g \boxtimes_{\mathcal C_2} M_{h} \cong M_{g\cdot h}$.
Let $L_0^1$ be the Virasoro zero-mode of $V_1$ and $L_0^2$ be the Virasoro zero-mode of $V_2$. Assume that for all $h$ in $H$ the twist operator $e^{2\pi i L_0^1\otimes L_0^2}$ acts as the identity on $J_h\otimes M_h$. Then 
\[
A := \bigoplus_{h\in H} J_h\otimes M_h
\]
is a super VOA extension of $V_1\otimes V_2$. Assume that every object of $\mathcal C_2$ is a subquotient of an iterated tensor product of simple objects in $\mathcal C_2$. Let $X$ in $\mathcal C_2$ and define $X_h:= M_h \boxtimes_{\mathcal C_2} X$. If the semi-simple part of $e^{2\pi i L_0^1\otimes L_0^2}$ acts as a scalar on 
\[
\mathcal F(V_1 \otimes X) \cong_{\mathcal C} \bigoplus_{h\in H}J_h \otimes X_h 
\]
then this induced object is a local $A$-module and thus a module of the VOA $A$. In other words knowledge of conformal dimensions tells us if objects are modules for the extended VOA. Moreover $\mathcal F(V_1 \otimes X) $ is simple/projective if and only if $X$ is simple/projective. 
Let $X$ and $Y$ be two objects of $\mathcal C_2$ then by Frobenius reciprocity
\begin{equation}\label{eq:rightadjoint}
\begin{split}
\text{Hom}_{\cat^A}\left(\mathcal F(V_1 \otimes X), \mathcal F(V_1 \otimes Y)\right) &\cong \text{Hom}_{\mathcal C}\left(\mathcal F(V_1 \otimes X), V_1 \otimes Y\right)\\
 &\cong \text{Hom}_{\mathcal C}\left(\  \bigoplus_{h\in H}J_h \otimes X_h, V_1 \otimes Y\right)\\
  &\cong  \text{Hom}_{\mathcal C_2}\left(X, Y\right).
\end{split}
\end{equation}
where we used that $\text{Hom}_{\mathcal C_1}(J_h, V_1) = \delta_{h, e}\mathbb C$. 
We thus see that $\mathcal F(V_1 \otimes X)$ and $\mathcal F(V_1 \otimes Y)$ are isomorphic as $A$-modules if and only if $X\cong Y$ as $\mathcal C_2$-modules. 
Let now $M$ in $\mathcal C_2$ and 
\begin{equation}\nonumber
\dots  \longrightarrow P^4_M\longrightarrow P^3_M\longrightarrow P^2_M\longrightarrow P^1_M\longrightarrow P^0_M \longrightarrow M \longrightarrow 0
\end{equation} 
be a projective resolution of $M$, then the corresponding projective resolution in $\mathcal C$ is 
\begin{equation}\nonumber
\dots  \longrightarrow V_1 \otimes P^4_M\longrightarrow V_1 \otimes P^3_M\longrightarrow V_1 \otimes P^2_M\longrightarrow V_1 \otimes P^1_M\longrightarrow V_1 \otimes P^0_M \longrightarrow V_1 \otimes M \longrightarrow 0
\end{equation} 
and via induction we get the projective resolution in $\mathcal C^A$
\begin{equation}\nonumber
\dots \longrightarrow  \mathcal F(V_1 \otimes P^3_M)\longrightarrow  \mathcal F(V_1 \otimes P^2_M)\longrightarrow  \mathcal F(V_1 \otimes P^1_M)\longrightarrow  \mathcal F(V_1 \otimes P^1_M) \longrightarrow  \mathcal F(V_1 \otimes M) \longrightarrow 0
\end{equation}
This is a projective resolution in $\catA$ provided $\mathcal F(V_1 \otimes M)$ is local. Assume now that indeed $\mathcal F(V_1 \otimes M)$ is local.
Using \eqref{eq:rightadjoint} we get that 
\[
\text{Hom}_{\catA}(\mathcal F(V_1 \otimes P^n_M), \mathcal F(X)) \cong \text{Hom}_{\mathcal C_2}(P^n_M, X)
\]
and thus the exact functor $\mathcal F$ maps Hom of the projective resolution of $M$ to Hom of the projective reslution of $\mathcal F(V_1\otimes M)$, i.e. 
\[
0 \longrightarrow \text{Hom}_{\mathcal C_2}(P^0_M, X) \longrightarrow \text{Hom}_{\mathcal C_2}(P^1_M, X) \longrightarrow \dots
\]
is mapped to
\begin{equation}\nonumber
\begin{split}
0 \longrightarrow \text{Hom}_{\catA}(\mathcal F(V_1 \otimes P^0_M), \mathcal F(X)) \longrightarrow \text{Hom}_{\catA}(\mathcal F(V_1 \otimes P^1_M), \mathcal F(X))\longrightarrow \dots 
\end{split}
\end{equation}
and especially cohomology rings are isomorphic
\[
\text{Ext}_{\mathcal C_2}^\bullet(M, M) \cong \text{Ext}_{\catA}^\bullet(\mathcal F(V_1\otimes M), \mathcal F(V_1\otimes M))
\]
and also their modules (for $ \mathcal F(V_1\otimes X)$ a local module)
\[
\text{Ext}_{\mathcal C_2}^\bullet(M, X) \cong \text{Ext}_{\catA}^\bullet(\mathcal F(V_1\otimes M), \mathcal F(V_1\otimes X)).
\]

\subsubsection{Example 2}\label{ex:two}

Let $V$ be a VOA and $\mathcal C$ a full rigid vertex tensor category of $V$-modules. Let $G$ be a group of simple currents $M_g$ such that
\[
B := \bigoplus_{g \in G} M_g
\]
is a super VOA extending $V$. Let $M, X$ be objects of $\mathcal C$, such that $\mathcal F(M), \mathcal F(X)$ are local $B$-modules and let 
\begin{equation}\nonumber
\dots  \longrightarrow P^4_M\longrightarrow P^3_M\longrightarrow P^2_M\longrightarrow P^1_M\longrightarrow P^0_M \longrightarrow M \longrightarrow 0
\end{equation} 
be a projective resolution of $M$ so that 
\begin{equation}\nonumber
\dots  \longrightarrow \mathcal F(P^4_M)\longrightarrow \mathcal F(P^3_M)\longrightarrow \mathcal F(P^2_M)\longrightarrow \mathcal F(P^1_M)\longrightarrow \mathcal F(P^0_M) \longrightarrow \mathcal F(M) \longrightarrow 0
\end{equation} 
is a projective resolution of $\mathcal F(M)$. Let $X_g:= M_g\boxtimes_{\mathcal C}X$, then 
\begin{equation}
\begin{split}
\text{Hom}_{\catB}(\mathcal F(P^n_M), \mathcal F(X)) &\cong \text{Hom}_{\mathcal C}(P^n_M, B\boxtimes_{\mathcal C} X) \\
&\cong  \bigoplus_{g\in G} \text{Hom}_{\mathcal C}(P^n_M, X_g)
\end{split}
\end{equation}
so that in this case comparing the cohomologies of the Hom-spaces of the projective resolutions tells us that
\[
\text{Ext}_{\mathcal C}^\bullet(M, B\boxtimes_{\mathcal C} M) \cong \text{Ext}_{\catB}^\bullet(\mathcal F(M), \mathcal F(M))
\]
and also 
\[
\text{Ext}_{\mathcal C_2}^\bullet(M, B\boxtimes_{\mathcal C}X) \cong \text{Ext}_{\catB}^\bullet(\mathcal F(M), \mathcal F(X)).
\]

The role of $V_1$ of the first example will be played by a tensor product of a lattice VOA and a Heisenberg VOA, while $V_2$ of the first example and the VOA $V$ of the second example will be given by (tensor products of) singlet VOAs. The triplet VOA will be realized in the spirit of example 2, the affine super VOA of $\mathfrak{psl}(N|N)$ at level one is of the type of example one and its simple current extension needs a combination of both examples. 

\subsubsection{Lifting logarithmic modules}\label{sec:liftlog}

We call a module logarithmic if the Virasoro zero-mode does not act semi-simple. Let $V_1, V_2$ be VOAs with rigid vertex tensor categories $\mathcal C_1, \mathcal C_2$. We don't require either of them to be locally finite, i.e. objects might very well have infinite Jordan-H\"older length. Let $\mathcal C=\mathcal C_1 \boxtimes \mathcal C_2$ and let 
\[
A  = \bigoplus_{g\in G} J_g \otimes K_g
\]
be a simple current extension for the abelian group $G$ generated by $J_1\otimes K_1, \dots, J_n\otimes K_n$. Let $N\otimes M$ be an object of $\mathcal C$. Then 
\[
\mathcal F(N\otimes M) \cong_{\mathcal C} A \boxtimes_{\mathcal C} (N\otimes M) 
\]
is a local $A$-module if and only if the monodromy, i.e. the double braiding,  
\[
M_{J_i\otimes K_i, N\otimes M} \in \text{End}((J_i\otimes K_i) \boxtimes_{\mathcal C} (N\otimes M))
\]
is trivial. Let $h_M$ denote the conformal weight of the top level of the module $M$. 
 The semi-simple part of the monodromy is just given by $e^{2\pi i \Delta_i}$ with $\Delta_i$ the sum 
\[
\Delta_i = h_{(J_i\otimes K_i) \boxtimes_{\mathcal C} (N\otimes M)} -  h_{J_i\otimes K_i} - h_{N\otimes M}
\]
so that a necessary condition for $\mathcal F(N\otimes M)$ being local is that $\Delta_i =0\mod 1$ for all $i=1, \dots, n$. 
As said before this is also a sufficient condition if $N\otimes M$ is a subquotient of an iterated tensor product of simples in $\mathcal C$. 
If not then we define the modules $X_i, Y_i$ iteratively as follows. Firstly,  $X_0:=N\otimes M$ and $X_i, Y_i$ are defined recursively as the co-equalizer of monodromy and  identity in the following sense:
\[
\xymatrixcolsep{3.0pc}
\hspace{-1cm}
\xymatrix{
 (J_i\otimes K_i) \boxtimes_{\mathcal C} X_{i-1} \ \ \  \ar@<-.5ex>[r]_{M_{J_i\otimes K_i, X_{i-1}}} \ar@<.5ex>[r]^{\text{Id}}  & \ \ \ (J_i\otimes K_i) \boxtimes_{\mathcal C} X_{i-1} \ar[r] &  Y_i   \ar[r] & 0
}
\]
and 
\[
X_i := (J_i\otimes K_i)^{-1} \boxtimes_{\mathcal C} Y_i.
\]
This procedure ensures that $X_n$ lifts to a local module of $A$ and we will see in examples that it will give rise to quite useful local modules for our purposes.

\subsection{Heisenberg and lattice VOAs}\label{sec:HeisenbergVOA}

The simplest example of a VOA is the Heisenberg VOA, which physicists call the free boson. Let $V$ be a finite-dimensional vector space (over $\mathbb C$ and we set $n:=\text{dim}\ V$) and $\kappa: V \times V \rightarrow \mathbb C$ a bilinear form on $V$. One usually requires the form to be non-degenerate. Then the Heisenberg VOA associated to the vector space $V$ is strongly generated by fields $v_i(z)$ associated to a basis $\{ v_i\}$ of $V$ with OPE
$v_i(z)v_j(w) = \kappa(v_i, v_j)(z-w)^{-2}$. The underlying Lie algebra of modes is 
\[
H(V) = V\otimes_{\mathbb C} \mathbb C[[t, t^{-1}]]  \oplus \mathbb C K \oplus \mathbb C d.
\]
We write $v_{i, n}$ for $v_i\otimes t^n$. The commutation relations are $[v_{i, n}, v_{j, m}] =  \kappa(v_i, v_j)K \delta_{n+m, 0}n$, $K$ is central and $d$ is a derivation.
A weight $\lambda$ is a linear map $\lambda : V\oplus \mathbb C K \oplus \mathbb C d \rightarrow \mathbb C$, i.e. it defines a one-dimensional representation $\mathbb C_\lambda$ of $V\oplus \mathbb C K \oplus \mathbb C d$ and 
the Fock module of weight $\lambda$ is the induced highest-weight module 
\[
\mathcal F_\lambda := \text{Ind}^{H(V)}_{H(V)_0\oplus H(V)_+} \mathbb C_\lambda.
\]
We denote the category whose objects are direct sums of Fock modules by $\catfock$. The subcategory of real weight modules is known to be a vertex tensor category \cite[Thm.2.3]{CKLR} and fusion rules of Fock modules are just
\[
\mathcal F_\lambda \boxtimes_{\catfock} \mathcal F_\mu \cong \mathcal F_{\lambda+\mu}.
\]
This means that every Fock module is a simple current, i.e. an invertible element of the tensor ring. 

One usually chooses those $\lambda$ for which $K$ acts as the identity and $d$ as zero. We now fix a basis $\{v_i\}$ of $V$, such that $\kappa(v_i, v_j)$ is integral for all $i, j$, so that $L=\mathbb Z v_1 \oplus \dots \oplus  \mathbb Z v_n$ is a integral lattice in $V$. To each element $x$ of $L$ one identifies a corresponding weight via the bilinear form $\lambda = \kappa( \ \ ,  x )$ and letting $\lambda(K)=1$ and $\lambda(d)=0$. We denote both the element of $L$ and the corresponding weight by $\lambda$. The lattice VOA $V_L$ is then the simple current extension
\[
V_L = \bigoplus_{\lambda\in L}\mathcal F_\lambda.
\]  
We are interested in two examples. Firstly, where $L=A_n$ is the root lattice of $\mathfrak{sl}(n+1)$ and secondly where $L=\sqrt{-1}A_n$.

The Heisenberg VOA gives us also a simple example that illustrates the behavior of Ext-groups. For this we enlarge the category of Fock-modules $\catfock$ by allowing for a non semi-simple action of the zero-modes $v_{i, 0}$. Then we have self-extensions of Fock modules that we denote by $\mathcal F^{(n)}_\lambda$ and the superscript indicates the Jordan-H\"older length of the extensions with each composition factor being isomorphic to the Fock module $\mathcal F_\lambda$. Let us call this category $\catfocklog$. While the category $\catfock$ is semisimple we have
\[
\text{Ext}^\bullet_{\catfocklog} (\mathcal F_\lambda, \mathcal F_\mu) \cong \delta_{\lambda, \mu} \mathbb C[x] 
\]
with $x^n$ corresponding to the extension
\[
0 \longrightarrow \mathcal F_\lambda \longrightarrow \mathcal F^{(2)}_\lambda \longrightarrow \mathcal F^{(2)}_\lambda \longrightarrow  \dots \longrightarrow \mathcal F^{(2)}_\lambda  \longrightarrow \mathcal F^{(2)}_\lambda  \longrightarrow \mathcal F_\lambda \longrightarrow 0
\]
in $\text{Ext}^n_{\catfocklog} (\mathcal F_\lambda, \mathcal F_\lambda)$.

\subsection{The singlet $M(p)$}\label{sec:singlet}

As reference we recommend \cite{CRW, CGR}.
Especially all relevant data is compactly summarized in section 5 of \cite{CGR}. Our main assumptions are the correctness of Conjecture 5.1 and Conjecture 5.8 of \cite{CGR}. These conjectures are tested in many ways in \cite{CGR, CM, CMR} and comparisons are given in section 5.4 of \cite{CGR}. The up-shot of correctness of these Conjectures is the knowledge of the complete vertex tensor category of finite-dimensional weight modules of the singlet VOA and especially this is a rigid and braided tensor category so that all the results of \cite{CKM} apply. We also would like to mention that the subtleties of passing to a completion of the category due to infinite order simple current extensions are discussed in both \cite{AR, CGR}.

The simple modules of the singlet VOA that we need are denoted $M_{s, k}$ with $k, s$ integer and $1\leq s \leq  p$. If $s=p$ then this module is projective and otherwise its projective cover has Loewy diagram as in Figure \ref{fig:Loewy_proj}. We also introduce the Zig-Zag modules as in Figure \ref{fig:Loewyzigzag}. The only fusion products that we need to know are the ones of the simple currents $M_{1, k}$ with simple and projective modules. They are 
\begin{equation}
M_{1, k} \boxtimes_{M(p)} M_{s, k'} \cong M_{s, k+k'-1}, \qquad  M_{1, k} \boxtimes_{M(p)} P_{s, k'} \cong P_{s, k+k'-1}.
\end{equation}
Let us first list all possible resolutions of projective modules. First, possible submodules and quotients are given by the following list: 
\begin{equation}
\begin{split}
&0 \longrightarrow  M_{s, k} \longrightarrow P_{s, k} \longrightarrow  \overline{Z}_{s, k} \longrightarrow 0 \\
&0 \longrightarrow  T^\pm_{s, k} \longrightarrow P_{s, k} \longrightarrow  T^\pm_{p-s, k \mp 1} \longrightarrow 0 \\
&0 \longrightarrow  Z_{s, k} \longrightarrow P_{s, k} \longrightarrow  M_{s, k} \longrightarrow 0 \\
\end{split}
\end{equation}
These submodules themselves decompose as
\begin{equation}
\begin{split}
a^\pm_{s, k}: \ \  &0 \longrightarrow  M_{s, k} \xrightarrow{e^\pm(t, s, k)} T^\pm_{s, k}\xrightarrow{p^\pm(t, s, k)}  M_{p-s, k\pm 1} \longrightarrow 0 \\
&0 \longrightarrow  M_{p-s, k-1} \oplus M_{p-s, k+1} \xrightarrow{e(\bar z, s, k)} \overline{Z}_{s, k} \longrightarrow  M_{s, k} \longrightarrow 0\\
&0 \longrightarrow  M_{p-s, k \pm 1} \longrightarrow \overline{Z}_{s, k} \longrightarrow  T^\pm_{p-s, k\mp 1} \longrightarrow 0\\
&0 \longrightarrow  M_{s, k}  \longrightarrow {Z}_{s, k} \xrightarrow{p(z, s, k)}  M_{p-s, k-1} \oplus M_{p-s, k+1} \longrightarrow 0\\
&0 \longrightarrow  T^\pm_{s, k}  \longrightarrow {Z}_{s, k} \xrightarrow{p^\pm(z, s, k)}  M_{p-s, k\mp1}\longrightarrow 0\\
\end{split}
\end{equation}
Moreover, the pullback of the projections $ p^\pm_{t, p-s, k \mp 1}: T^\pm_{p-s, k\mp 1} \rightarrow M_{s, k}$ is $p^*(M_{s, k})=\overline{Z}_{s, k}$  and the pushout of the embeddings $e^\pm_{t, s, k}: M_{s, k} \rightarrow T^\pm_{s, k}$ is $e_*(M_{s, k})=Z_{s, k}$. The Baer sum $+_{Baer}$ of two elements in  $\text{Ext}^n(A, B)$ is defined as
\begin{equation}
\begin{split} 
&0  \longrightarrow B \xrightarrow{\ \ \iota_1\ \ } X_n \longrightarrow X_{n-1} \longrightarrow \dots \longrightarrow X_{1} \xrightarrow{\ \ \pi_1\ \ } A \longrightarrow 0 \quad +_{Baer} \\
&0  \longrightarrow B \xrightarrow{\ \ \iota_2\ \ } Y_n \longrightarrow Y_{n-1} \longrightarrow \dots \longrightarrow Y_{1} \xrightarrow{\ \ \pi_2\ \ } A \longrightarrow 0 \quad = \\
&0  \rightarrow B \rightarrow  \iota_*(B) \rightarrow X_{n-1}\oplus Y_{n-1} \rightarrow \dots \rightarrow X_{2} \oplus Y_2  \rightarrow \pi^*(A) \rightarrow A \rightarrow 0.
 \end{split}
\end{equation}
with $\iota_*$ the pushout of $\iota_{1, 2}$ and $\pi^*$ the pullback of $\pi_{1, 1}$.
We thus see, that 
\begin{equation}
\begin{split}
&0 \longrightarrow M_{s, k} \longrightarrow T^+_{s, k} \longrightarrow T^-_{p-s, k+1} \longrightarrow M_{s, k} \longrightarrow 0 \quad +_{Baer} \\
&0 \longrightarrow M_{s, k} \longrightarrow T^-_{s, k} \longrightarrow T^+_{p-s, k-1} \longrightarrow M_{s, k} \longrightarrow 0 \quad = \\
&\qquad \qquad 0 \longrightarrow M_{s, k} \longrightarrow Z_{s, k} \longrightarrow \overline{Z}_{s, k} \longrightarrow M_{s, k} \longrightarrow 0.
 \end{split}
\end{equation}

Exact sequences are equivalent if they are related by a chain complex. 
We have for example the chain complex
\begin{equation}\label{eq:relyoneda}
 \xymatrixcolsep{2.5pc}
 \xymatrix{
 & 0 \ar[r] & M_{s, k} \ar[d]_{=} \ar[r]^{e \ \ \ } & T^-_{s, k} \oplus T^+_{s, k} \ar[d]_{p_\pm} \ar[r] & P_{s, k} \ar[d] \ar[r]  & M_{s, k} \ar[d]_{=} \ar[r] & 0 \\  
 & 0 \ar[r] & M_{s, k}  \ar[r] & T^\pm_{s, k} \ar[r] & T^\mp_{p-s, k \pm 1}  \ar[r]  & M_{s, k}  \ar[r] & 0 \\ 
 }
\end{equation}
Here, the second map $p_\pm$ is the projecton on the corresponding summand $T^\pm_{s, k}$ and $e=\left(e^-(t, s, k), e^+(t, s, k) \right)$.
The remaining maps are non-zero and as such all unique up to isomorphism. This gives us an imortant relation:
\begin{equation}\label{eq:relyoneda2}
[a^{+}_{s, k}] \circ [a^-_{p-s, k-1}] =   [a^{-}_{s, k}] \circ [a^+_{p-s, k+1}]
\end{equation}
here $[ \ \ ]$ denotes the equivalence class of the exact sequence and $\circ$ the class obtained by splicing the exact sequences (the bottom row of \eqref{eq:relyoneda}). 

We also have 
\begin{equation}
 \xymatrixcolsep{2.5pc}
 \xymatrix{
 & 0 \ar[r] & M_{s, k} \ar[d]_{=} \ar[r]^{\ \ \ } & T^-_{s, k} \oplus T^+_{s, k} \ar[d]_{} \ar[r] & P_{s, k} \ar[d] \ar[r]  & M_{s, k} \ar[d]_{=} \ar[r] & 0 \\  
 & 0 \ar[r] & M_{s, k}  \ar[r] & Z_{s, k} \ar[r] & \overline{Z}_{s, k}  \ar[r]  & M_{s, k}  \ar[r] & 0 \\ 
 }
\end{equation}
where the morphisms should be clear. 

\begin{figure}[tb]
\begin{center}
\begin{tikzpicture}[scale=0.70][thick,>=latex,
nom/.style={circle,draw=black!20,fill=black!20,inner sep=1pt}
]
\node(middle0) at  (14,-1) [] {$\dots$}; 
\node (up1) at (2.5,0) [] {$M_{p-s, k+n}$};
\node (up2) at (7.5,0) [] {$M_{p-s, k+n-2}$};
\node (dup2) at (12.5,0) [] {};
\node (up3) at (17.5,0) [] {$M_{p-s, k-n+2}$};
\node (up4) at (22.55,0) [] {$M_{p-s, k-n}$};
\node (bot1) at (5,-2.5) [] {$M_{s, k+n-1}$};
\node (bot2) at (10,-2.5) [] {$M_{s, n-3}$};
\node (dbot2) at (15,-2.5) [] {};
\node (bot3) at (20,-2.5) [] {$M_{s, k-n+1}$};
\draw [->] (up1) -- (bot1);
\draw [->] (up2) -- (bot1);
\draw[->] (up2) -- (bot2);
\draw[dashed,->] (dup2) -- (bot2);
\draw[dashed,->] (up3) -- (dbot2);
\draw [->] (up3) -- (bot3);
\draw [->] (up4) -- (bot3);

\end{tikzpicture}
\captionbox{\label{fig:Loewyzigzag} The Loewy diagram of the Zig-Zag $Z^n_{s,k}$.}{\rule{12cm}{0cm}}
\end{center}
\end{figure}

We now proceed in searching projective resolutions of the simple modules $M_{s, k}$.
Firstly, we realize Zig-Zag modules as images and kernels
\begin{equation}\nonumber
0 \longrightarrow  \bigoplus_{\substack{i=0 \\ \text{step}  2 }}^{2n-2}M_{s, k+n-1-i} \xrightarrow{e^n(\bar z, s, k)} T^+_{p-s, k+n-1} \oplus  T^-_{p-s, k-n+1} \oplus\bigoplus_{\substack{i=2 \\ \text{step}  2 }}^{2n-2} \overline{Z}_{p-s, k+n-i} \longrightarrow  {Z}^{n}_{s, k} \longrightarrow 0 
\end{equation}
where
\begin{equation}\nonumber
e^n(\bar z, s, k)\Big\vert_{M_{s, k+n-1-i}} = \begin{cases}  e(\bar z, p-s, k+n-2-i) \oplus e(\bar z, p-s, k+n-i) & \text{otherwise} \\ e^+(t, s, k+n-1) \oplus e(\bar z, p-s, k+n-2)  & \text{if} \ i = 0 \\
e^-(t, s, k-n+1) \oplus e(\bar z, p-s, k-n+2)  & \text{if} \ i = 2n-2 \\
\end{cases}
\end{equation}
$e^n(\bar z, s, k)$ restricted to the $i$-th summand is just $e(\bar z, p-s, k+n-i2)$ except for the first two it is $e^\pm(t, p-s, k\mp n\pm 1)$.
Note that the Zig-Zag module has as submodules  $\overline{Z}_{p-s, k+n-2i}$ for $i=0, \dots, n$ as well as $T^\pm_{s,k\pm(n-1)}$. 
Secondly, we have 
\begin{equation}\nonumber
0 \longrightarrow  {Z}^{n}_{s, k}  \longrightarrow T^+_{s, k+n-2} \oplus T^-_{k-n+2} \oplus \bigoplus_{\substack{i=2 \\ \text{step}  2 }}^{2n-4} \overline{Z}_{p-s, k+n-1-i} \xrightarrow{p^n(z, s, k)}  \bigoplus_{\substack{i=0 \\ \text{step}  2 }}^{2n-4}M_{s, k+n-2-i} \longrightarrow 0 
\end{equation}
where $p^n(z, s, k)$ restricted to the $i$-th summand is $p(z, p-s, k+n-1-2i)$ and for the first two it is $p^\pm(t, s, k\pm n\mp 2)$.
\begin{figure}[tb]
\begin{center}
\begin{tikzpicture}[scale=0.70][thick,>=latex,
nom/.style={circle,draw=black!20,fill=black!20,inner sep=1pt}
]
\node(left0) at  (-1,0) [] {$P_{s,k}$:}; 
\node (top1) at (5,2.5) [] {$M_{s, k}$};
\node (left1) at (2.5,0) [] {$M_{p-s, k+1}$};
\node (right1) at (7.5,0) [] {$M_{p-s, k-1}$};
\node (bot1) at (5,-2.5) [] {$M_{s, k}$};
\draw [->] (top1) -- (left1);
\draw [->] (top1) -- (right1);
\draw [->] (left1) -- (bot1);
\draw [->] (right1) -- (bot1);
\end{tikzpicture}
\captionbox{\label{fig:Loewy_proj} The Loewy diagram of the projective cover  $P_{s,k}$  of the simple module $M_{s,k}$.}{\rule{12cm}{0cm}}
\end{center}
\end{figure}
The projective cover of a Zig-Zag is a sum of indecomposable projectives, i.e.
\begin{equation}
0\longrightarrow Z^{n+1}_{p-s, k} \longrightarrow \bigoplus_{\substack{ \ell=-n \\ \ell \ \text{even}}}^n P_{p-s,k+\ell } \longrightarrow Z^n_{s, k} \rightarrow 0
\end{equation}
Define
\begin{equation}
P^n_{s, k} = \bigoplus_{\substack{ \ell=-n \\ \ell \ \text{even}}}^n P_{s,k+\ell }
\end{equation}
so that splicing this series of short-exact sequences yields the projective resolution
\begin{equation}\label{eq:projsing}
\dots  \longrightarrow P^5_{p-s, k}\longrightarrow P^4_{s, k}\longrightarrow P^3_{p-s, k}\longrightarrow P^2_{s, k}\longrightarrow P^1_{p-s, k}\longrightarrow P^0_{s, k} \longrightarrow M_{s, k} \longrightarrow 0.
\end{equation} 
Taking Hom of it we have 
\begin{equation}\nonumber
0 \longrightarrow  \text{Hom}(P_{s, k}^0,  \bullet \, )  \longrightarrow  \text{Hom}(P_{p-s, k}^1,   \bullet \, )  \longrightarrow  \text{Hom}(P_{s, k}^2,   \bullet \, )  \longrightarrow  \text{Hom}(P_{p-s, k}^3,   \bullet \, ) \dots 
\end{equation}
with 
\begin{equation}
 \text{Hom}(P_{t, k}^m,  M_{s, k} ) \cong \begin{cases} \mathbb C & \text{if} \ m \ \text{is} \ \text{even and } \ t=s \\  \mathbb C & \text{if} \ m \ \text{is} \ \text{odd and }\ t=p-s\\
 0 & \text{else} \end{cases} 
\end{equation}
and so we especially have
\[
\text{Ext}_{\mathcal C}^\bullet(M_{s, k}, M_{s, k}) \cong \bigoplus_{\substack{m=0 \\ m \ \text{even} }}^\infty \mathbb C \cong \mathbb C[x^2]
\]
as Ext-algebra with $x^2:= [a^\pm_{s, k}] \circ [a^\mp_{p-s, k\pm 1}]$ and $\circ$ is splicing of exact sequences. 

\subsubsection{Extension in $\mathcal C_{\text{log}}$}\label{sec:Clog}

In \cite{CMR} a larger category of quantum group modules was studied and conjectured to be equivalent as braided tensor category to the category $\catlog$  of finite length modules for the singlet algebra. This category has not been studied much.
The important new ingredient is that in this category we allow weight spaces to be generalized eigenspaces for $H$. Here $H$ is the zero-mode of the Heisenberg VOA of which the singlet $M(p)$ is a sub VOA. One especially expects that in this category the modules $P_{s, k}$ allow for self-extensions so that modules with Loewy diagram as in Figure \ref{fig:logproj} and \ref{fig:logprojR} appear as quotient modules. 
\begin{figure}[tb]
\begin{center}
\begin{tikzpicture}[scale=0.70][thick,>=latex,
nom/.style={circle,draw=black!20,fill=black!20,inner sep=1pt}
]
\node(left0) at  (0,0) [] {$P^L_{s,k}$:}; 
\node (top1) at (5,2.5) [] {$M_{s, k}$};
\node (left1) at (2.5,0) [] {$M_{p-s, k+1}$};
\node (right1) at (7.5,0) [] {$M_{p-s, k-1}$};
\node (bot1) at (5,-2.5) [] {$M_{s, k}$};
\draw [->] (top1) -- (left1);
\draw [->] (top1) -- (right1);
\draw [dashed, ->] (top1) -- (bot1);
\draw [->] (right1) -- (bot1);

\node(Left0) at  (10,0) [] {$P^R_{s,k}$:}; 
\node (Top1) at (15,2.5) [] {$M_{s, k}$};
\node (Left1) at (12.5,0) [] {$M_{p-s, k+1}$};
\node (Right1) at (17.5,0) [] {$M_{p-s, k-1}$};
\node (Bot1) at (15,-2.5) [] {$M_{s, k}$};
\draw [->] (Top1) -- (Left1);
\draw [->] (Top1) -- (Right1);
\draw [->] (Left1) -- (Bot1);
\draw [dashed, ->] (Top1) -- (Bot1);
\end{tikzpicture}
\captionbox{\label{fig:logproj} New modules extending  the simple module $M_{s,k}$ in $\mathcal C_{\text{log}}$. The dashed-line indicates the nilpotent action of $H$.}{\rule{12cm}{0cm}}
\end{center}
\end{figure}
\begin{figure}[tb]
\begin{center}
\begin{tikzpicture}[scale=0.70][thick,>=latex,
nom/.style={circle,draw=black!20,fill=black!20,inner sep=1pt}
]
\node(left0) at  (0,0) [] {$R_{s,k}$:}; 
\node (top1) at (5,2.5) [] {$M_{s, k}$};
\node (left1) at (2.5,0) [] {$M_{p-s, k+1}$};
\node (right1) at (7.5,0) [] {$M_{p-s, k-1}$};
\node (bot1) at (5,-2.5) [] {$M_{s, k} \oplus M_{s, k}$};
\draw [->] (top1) -- (left1);
\draw [->] (top1) -- (right1);
\draw [dashed, ->] (top1) -- (bot1);
\draw [->] (right1) -- (bot1);
\draw [->] (left1) -- (bot1);
\end{tikzpicture}
\captionbox{\label{fig:logprojR} New modules extending  the simple module $M_{s,k}$ in $\mathcal C_{\text{log}}$. The dashed-line indicates the nilpotent action of $H$.}{\rule{12cm}{0cm}}
\end{center}
\end{figure}

One gets then the following chain complex
\begin{equation}\label{eq:relyonedalog}
 \xymatrixcolsep{2.0pc}
\hspace{-1cm} \xymatrix{
 & 0 \ar[r] & M_{s, k} \ar[d]_{=} \ar[r] & T^+_{s, k}  \ar[d]_{p_\pm} \ar[r] & T^-_{p-s, k+1} \ar[d] \ar[r]  & M_{s, k} \ar[d]_{=} \ar[r] & 0 \\  
 & 0 \ar[r] & M_{s, k}  \ar[r] & P_{s, k}^R \ar[r] & T^+_{p-s, k - 1} \oplus T^-_{p-s, k + 1} \ar[r]  & M_{s, k}  \ar[r] & 0 \\ 
 & 0 \ar[r] & M_{s, k}\ar[u]_{=}  \ar[r] & M_{s, k} \oplus M_{p-s, k-1} \ar[u]_{}\ar[r] & T^+_{p-s, k - 1}  \ar[u]_{}\ar[r]  & M_{s, k} \ar[u]_{=} \ar[r] & 0 \\ 
 }
\end{equation}
where all morphisms should be clear as for each map there is exactly one non-trivial possibility. It thus follows that in $\mathcal C_{\text{log}}$ the element $x^2:= [a^\pm_{s, k}] \circ [a^\mp_{p-s, k\pm 1}]$ of $\text{Ext}^2(M_{r, s}, M_{r, s})$ is equivalent to a split exact sequence and thus trivial. It follows that 
\[
\text{Ext}_{\mathcal C_{\text{log}}}^\bullet(M_{s, k}, M_{s, k}) \cong \mathbb C.
\]

We now turn to multiple copies of $\mathcal C_{\text{log}}$.

\subsubsection{Many copies of $\mathcal C_{\text{log}}$}\label{sec:manyClog}

Let us consider $N$ copies of $\mathcal C_{\text{log}}$, i.e. the $n$-fold Deligne product of this category. 
However we will require that the nilpotent parts $H_i^{\text{nil}}$ to satisfy 
\[
\sum_{i=1}^N H_i^{\text{nil}}  = 0.
\]
We denote this category by $\mathcal C^N_{\text{log}}$.  
Let us take $N=2$ for the moment and we will see that the general case follows via the obvious embeddings $e_{i, j}$ of $\mathcal C^2_{\text{log}}$ in $\mathcal C^N_{\text{log}}$ into the $i$-th and $j$-th factor:
\[
e_{i, j}: \mathcal C^2_{\text{log}} \mapsto \bf 1 \boxtimes \dots \boxtimes \bf 1 \boxtimes   \mathcal C_{\text{log}} \boxtimes \bf 1 \boxtimes \dots \boxtimes \bf 1 \boxtimes   \mathcal C_{\text{log}} \boxtimes \bf 1 \boxtimes \dots \boxtimes \bf 1 \subset \mathcal C^N_{\text{log}}.
\]
The relevant modules are introduced via their Loewy diagrams in the following Figure \ref{fig:logproj2}. They are defined as follows, let $Q_{s, k, s', k'} $ be the quotient
\[
0\longrightarrow \left(T_{s, k}^+ \oplus M_{p-s, k-1}\right) \otimes \left(T_{s', k'}^+ \oplus M_{p-s', k'-1}\right)
\longrightarrow P^R_{s, k} \otimes P^R_{s', k'} \longrightarrow Q_{s, k, s', k'} \longrightarrow 0
\]
and let $X_{s, k, s', k'}$ be the coequalizer of the nilpotent parts of $H_1$ and $-H_2$
\[
\xymatrixcolsep{3.0pc}
\hspace{-1cm}
\xymatrix{
 Q_{s, k, s', k'}  \ar@<-.5ex>[r]_{H_1^{\text{nil}}} \ar@<.5ex>[r]^{-H_2^{\text{nil}}}  & Q_{s, k, s', k'} \ar[r] &  X_{s, k, s', k'}   \ar[r] & 0.
}
\]
\begin{figure}[tb]
\begin{center}
\begin{tikzpicture}[scale=0.70][thick,>=latex,
nom/.style={circle,draw=black!20,fill=black!20,inner sep=1pt}
]
\node(left0) at  (0,3) [] {$X_{s,k, s', k'}$:}; 
\node (top1) at (9,4) [] {$M_{s, k}\otimes M_{s', k'}$};
\node (left1) at (0,0) [] {$M_{p-s, k+1}\otimes M_{s', k'}$};
\node (left2) at (5,0) [] {$M_{s, k}\otimes M_{p-s', k'+1}$};
\node (right1) at (13,0) [] {$M_{s, k}\otimes M_{p-s', k'-1}$};
\node (right2) at (18,0) [] {$M_{p-s, k-1}\otimes M_{s', k'}$};
\node (bot1) at (9,-4) [] {$M_{s, k}\otimes M_{s', k'}$};
\draw [->] (top1) -- (left1);
\draw [->] (top1) -- (left2);
\draw [->] (top1) -- (right1);
\draw [->] (top1) -- (right2);
\draw [dashed, ->] (top1) -- (bot1);
\draw [->] (left1) -- (bot1);
\draw [->] (left2) -- (bot1);

\end{tikzpicture}
\captionbox{\label{fig:logproj2} New module $X_{s,k, s', k'}$ extending  the simple module $M_{s, k}\otimes M_{s', k'}$ in $\mathcal C_{\text{log}}\boxtimes \mathcal C_{\text{log}}$. The dashed-line indicates the nilpotent action of $H$.}{\rule{12cm}{0cm}}
\end{center}
\end{figure}
We also need the pushout $\iota_\pm^*(M_{s, k} \otimes M_{s', k'})$ of the embeddings $M_{s, k} \otimes M_{s', k'} \rightarrow T^\pm_{s, k} \otimes M_{s', k'}$ and $M_{s, k} \otimes M_{s', k'} \rightarrow M_{s, k} \otimes T^\pm_{s', k'}$ and the pullback $\pi^*_\pm(M_{s, k} \otimes M_{s', k'})$ of the surjections $T^\pm_{p-s, k\mp 1} \otimes M_{s', k'} \rightarrow M_{s, k} \otimes M_{s', k'}$ and $M_{s, k} \otimes T^\pm_{p-s', k\mp 1} \rightarrow M_{s, k} \otimes M_{s', k'}$. With this notation the Baer sum $x_1^2-x_2^2$ is just
\begin{equation}\nonumber
0 \longrightarrow M_{s, k} \otimes M_{s', k'} \longrightarrow \iota_\pm^*(M_{s, k} \otimes M_{s', k'}) \longrightarrow \pi^*_\mp(M_{s, k} \otimes M_{s', k'}) \longrightarrow M_{s, k} \otimes M_{s', k'} \longrightarrow 0.
\end{equation} 
The minus sign is needed so that we have maps from $\iota_\pm^*(M_{s, k} \otimes M_{s', k'})$ to $X_{s, k, s', k'}$ so that we get the chain complex
\begin{equation}\label{eq:relyonedalog}
 \xymatrixcolsep{2.0pc}
\hspace{-1cm} \xymatrix{
& 0 \ar[d] & 0 \ar[d] & 0 \ar[d] \\ 
& M_{s, k} \otimes M_{s', k'}  \ar[d] \ar[r]^{=} & M_{s, k} \otimes M_{s', k'}  \ar[d] & M_{s, k} \otimes M_{s', k'}  \ar[l]_{=}\ar[d]\\
& Y_{s, k, s', k'}   \ar[d]\ar[r]  & X_{s, k, s', k'}\ar[d] & \iota_+^*(M_{s, k} \otimes M_{s', k'}) \ar[l]\ar[d]   \\
& \pi^*_+(M_{s, k} \otimes M_{s', k'})\ar[d]\ar[r] &\pi^*_+(M_{s, k} \otimes M_{s', k'}) \oplus \pi^*_-(M_{s, k} \otimes M_{s', k'})\ar[d]& \pi^*_-(M_{s, k} \otimes M_{s', k'}) \ar[l]\ar[d] \\
& M_{s, k} \otimes M_{s', k'}  \ar[d] \ar[r]^{=} & M_{s, k} \otimes M_{s', k'}  \ar[d] & M_{s, k} \otimes M_{s', k'}  \ar[l]_{=}\ar[d]\\
& 0  & 0  & 0  \\ 
 }
\end{equation}
with the direct sum of simples
\[
Y_{s, k, s', k'} =M_{s, k} \otimes M_{s', k'} \oplus M_{s, k} \otimes M_{p-s', k'-1} \oplus M_{p-s, k-1} \otimes M_{s', k'} .
\]
We see that  $x_1^2=x_2^2$ in this category.
This obviously generalizes to $N$-copies of  $\mathcal C_{\text{log}}$ and then we have the relation 
\[
x_i^2=x_j^2\qquad \forall \ i, j. 
\]

\subsection{Many copies of $M(p)$ and a diagonal simple current extension}\label{sec:manysinglets}

We now realize the example of subsection \ref{ex:two}.
Let's take $N$ copies of the singlet VOA, then we can take the $N$-complex of the products of the projective resolutions and then restrict to the projective subcomplex and take its total complex. 
For this introduce $S=(s_1, \dots, s_N)$, $K=(k_1, \dots, k_N)$ and define 
\[
M_{S, K} = M_{s_1, k_1} \otimes  M_{s_2, k_2} \otimes  \dots  M_{s_{N-1}, k_{N-1}} \otimes  M_{s_N, k_N} 
\]
and 
\[
P^n_{S, K} = \bigoplus_{a_1+ \dots + a_N=n} P^{a_1}_{g(s_1, a_1), k_1} \otimes P^{(a_2}_{g(s_2, a_2), k_2} \otimes \dots \otimes P^{a_{N-1}}_{g(s_{N-1}, a_{N-1}), k_{N-1}} \otimes P^{a_N}_{g(s_N,a_n), k_N} 
\]
with $g(s, m)= s$ if $m$ is even and $p-s$ if $m$ is odd. 
so that the total complex is
\begin{equation}
\dots  \longrightarrow P^5_{S, K}\longrightarrow P^4_{S, K}\longrightarrow P^3_{S, K} \longrightarrow P^2_{S, K}\longrightarrow P^1_{S, K}\longrightarrow P^0_{S, K}
\end{equation} 
and this extends to the projective resolution
\begin{equation}
\dots  \longrightarrow P^5_{S, K}\longrightarrow P^4_{S, K}\longrightarrow P^3_{S, K} \longrightarrow P^2_{S, K}\longrightarrow P^1_{S, K}\longrightarrow P^0_{S, K} \longrightarrow M_{S, K} \longrightarrow 0
\end{equation} 
since the image of the map $P^1_{S, K}\longrightarrow P^0_{S, K}$ has the top $M_{S, K}$ as quotient. 
Taking Hom of it we have 
\begin{equation}
0 \longrightarrow  \text{Hom}(P_{S, K}^0,  \ \bullet \ )  \longrightarrow  \text{Hom}(P_{S, K}^1,  \ \bullet \ )  \longrightarrow  \text{Hom}(P_{S, K}^2,  \ \bullet \ )  \longrightarrow  \text{Hom}(P_{S, K}^3,  \ \bullet \ ) \dots 
\end{equation}
with 
\begin{equation}
 \text{Hom}(P_{S, K}^m,  M_{S, K} ) \cong \begin{cases} \text{Sym}_{\frac{m}{2}}\mathbb C^N & \text{if} \ m \ \text{is} \ \text{even} \\  0 & \text{if} \ m \ \text{is} \ \text{odd} \end{cases} 
\end{equation}
and so we have
\[
\text{Ext}_{\mathcal C}^\bullet(M_{S, K}, M_{S, K}) \cong \bigoplus_{\substack{m=0 \\ m \ \text{even} }}^\infty \text{Sym}_{\frac{m}{2}}\mathbb C^N \cong \mathbb C[x_1^2, \dots, x_N^2]
\]
and 
\[
\text{Ext}_{\mathcal C^N_{\text{log}}}^\bullet(M_{S, K}, M_{S, K})\cong \mathbb C[x_1^2, \dots, x_N^2]/(x_i^2=x_j^2) \cong \mathbb C[x^2]
\]
as Ext-algebras.

We are also interested in the diagonal simple current extension which then identifies modules accordingly. For this we in addition require $p=2$ (only necessary for odd $N$ so can be phrazed more general if desired). 
Then we restrict attention to singlet modules of type $M_{1, k}$, i.e. we fix the $s$-label to one ($s=2$ would also be possible but leads to projective modules that are not interesting for the present discussion). Let's denote the vector $\rho=(1, 1, \dots, 1)$. 
We are interested in the extensions
\begin{equation}\label{eq:algA}
A = \bigoplus_{k\in \mathbb Z}  M_{\rho, k\rho} 
\end{equation}
so that singlet modules $M_{\rho, K_1}, M_{\rho, K_s}$ lift to the same extended VOA module if and only if $K_1 = K_2 \mod  \mathbb Z\rho$, i.e. 
\[
\mathcal F(M_{\rho, K_1}) \cong \mathcal F(M_{\rho, K_2})\qquad \text{if and only if} \qquad
K_1 = K_2 \mod  \mathbb Z\rho.
\]
Here we denote the induction functor to local $A$-modules by $\mathcal F$. 
Then we have the additional extensions given by the images under the induction functor of 
\[
v_\pm^N = [a^\pm_{1, 1 \pm 1}] \otimes \dots \otimes [a^\pm_{1, 1\pm 1}]
\]
which clearly satisfy (recall \eqref{eq:relyoneda2}) 
\[
v_+^N \circ v_-^N = x_1^2 \otimes \dots \otimes x_N^2
\]
so that we obtain the Ext-algebra
\[
\text{Ext}_{\mathcal C}^\bullet(\mathcal F(M_{S, K}), \mathcal F(M_{S, K})) \cong \mathbb C[x_1^2, \dots, x_N^2, v_\pm^N]/ v_+^N  v_-^N - x_1^2 \cdot \dots \cdot x_N^2
\]
and 
\[
\text{Ext}_{\mathcal C^N_{\text{log}}}^\bullet(\mathcal F(M_{S, K}), \mathcal F(M_{S, K})) \cong \mathbb C[x^2, v_\pm^N]/v_+^Nv_-^N-x^{2N}
\]

\subsection{The triplet $W(p)$}\label{sec:triplet}

The triplet VOA is an extension of the singlet, 
\[
W(p) = \bigoplus_{k\in \mathbb Z} M_{1, 2k+1}
\]
and the simple triplet modules are the induced modules
\[
W_{s, r} = \mathcal F(M_{s, r+2k})
\]
with $\mathcal F$ the usual induction functor. The induction of the projectives is 
\[
R^n_{s, r} = \mathcal F(P^n_{s, r+2k}) \cong (n+1)R_{s, r+n}.
\]
Here, the $(n+1)$ could  be thought as the $n+1$ dimensional representation of $SU(2)$, $\rho_{n\omega}$, i.e. the projective resolution for triplet modules is  
\begin{equation}\label{eq:projtrip}
\dots \longrightarrow \rho_{3\omega} \otimes R_{p-s, 3-r} \longrightarrow \rho_{2\omega} \otimes R_{s, r}\longrightarrow \rho_{\omega} \otimes R_{p-s, 3-r} \longrightarrow \rho_{0} \otimes R_{s, r} \longrightarrow W_{s, r} \longrightarrow 0.
\end{equation} 
We see that 
\begin{equation}
\text{Ext}_{\mathcal C}(W_{s, r}, W_{s, r}) \cong \mathbb C[x, y]^{\mathbb Z_2}
\end{equation}
(with $\mathbb Z/2\mathbb Z =: \mathbb Z_2$) that is
\begin{equation}
\text{Spec}\left( \text{Ext}_{\mathcal C}(W_{s, r}, W_{s, r})\right)  = \mathbb C^2/\mathbb Z_2
\end{equation}
as expected. We remark that only objects of $\mathcal C\subset \mathcal C_{\text{log}}$ lift to local triplet VOA modules. 
The situation is different for the $\beta\gamma$-VOA as we will discuss in a moment.

\subsubsection{VOA extensions of $W(p)$}\label{sec:symplecticfermionext}

The triplet $W(p)$ allows for further VOA extension. The reason is that $W_{1, 2}$ is an order two simple current of quantum dimension $(-1)^{p+1}$ and twist $e^{2\pi i \left(\frac{3p-2}{4}\right)}$, see section 4 of \cite{CKL}. This means that for even $p$
\[
F(p) = \bigoplus_{k\in \mathbb Z} M_{1, k} = W_{1, 1} \oplus W_{1, 2}
\]
is another VOA extension. It is a $\mathbb Z$-graded super VOA if $p=2\mod 4$ and a $\frac{1}{2}\mathbb Z$-graded VOA if $p=0\mod 4$. We note that the case $p=2$ is the well-known sympectic fermion VOA. 
The induction functor $\mathcal F$ induces triplet modules as
\[
F_s := \mathcal F(W_{s, r}) \cong W_{s, 1} \oplus W_{s, 2}. 
\]
Where one checks that for odd $s$ the induced modules are local and for even $s$ twisted modules. The projective resolution \eqref{eq:projtrip} becomes via induction the projective resolution of $F(p)$-modules
\begin{equation}\label{eq:projsupertrip}
\dots \longrightarrow \rho_{3\omega} \otimes S_{p-s} \longrightarrow \rho_{2\omega} \otimes S_{s}\longrightarrow \rho_{\omega} \otimes S_{p-s} \longrightarrow \rho_{0} \otimes S_{s} \longrightarrow F_{s} \longrightarrow 0.
\end{equation} 
with 
\[
S_s := \mathcal F(R_{s, r}) \cong R_{s, 1} \oplus R_{s, 2}. 
\]
We thus see that 
\begin{equation}
\text{Ext}(F_{s}, F_{s}) \cong \begin{cases} \mathbb C[x, y] & \text{if} \ p=2 \\ \mathbb C[x, y]^{\mathbb Z_2} & \text{else} \end{cases}
\end{equation}
that is in the symplectic fermion case one has
\begin{equation}
\text{Spec}\left( \text{Ext}(F_{s}, F_{s})\right)  = \mathbb C^2
\end{equation}

\subsubsection{Orbifolds of $W(p)$}

Consider now the orbifold VOA $W(p)^{\mathbb Z_n}$. Let $g$ be a generator of $\mathbb Z_n=\mathbb Z/n\mathbb Z$. Then the action of $g$ is defined as $e^{2\pi i \frac{k}{n}}$ on the singlet submodule $M_{1, 2k+1}$. it thus follows that 
\[
W(p)^{\mathbb Z_n} = \bigoplus_{k\in \mathbb Z} M_{1, 2nk+1}
\]
We thus see that projective singlet modules $P_{r, s}$ and $P_{r', s'}$ lift to isomorphic $W(p)^{\mathbb Z_n}$-modules if and only if $r=r'$ and $s=s'\mod 2n$. From the projective resolution of singlet modules \eqref{eq:projsing} one thus sees using induction that 
\begin{equation}
\text{Ext}(\mathcal F(M_{s, r}), \mathcal F(M_{s, r})) \cong \mathbb C[x, y]^{\mathbb Z_{2n}}
\end{equation}
as expected.

\subsection{The $\beta\gamma$-VOA}\label{sec:betagammaext}

The $\beta\gamma$-VOA is denoted by $\mathcal B_2$ in \cite{CRW} and since we use the construction of that work we will also use the notation. It is a simple current extension of $M(2)\otimes \mathcal H$ and as such it is
\[
\mathcal B_2 \cong \bigoplus_{k\in \mathbb Z} M_{1, k+1} \otimes \mathcal F_{\sqrt{-1}k}.
\]
We allow for self-extensions of Fock modules $\mathcal F_\lambda$, for example $\mathcal F^{(2)}_\lambda$ denotes one self-extension, i.e.
\[
0\rightarrow \mathcal F_\lambda \rightarrow \mathcal F^{(2)}_\lambda \rightarrow \mathcal F_\lambda \rightarrow 0.
\]
We would like to lift the module $P^R_{1, 1} \otimes \mathcal F^{(2)}_0$ to a lcoal $\mathcal B_2$-module and so we have to apply the strategy outlined in section \ref{sec:liftlog}. The nilpotent part of the monodromies
\[
M_{\mathcal F^{(2)}_0, \mathcal F_{\sqrt{-1}}} \qquad \text{and} \qquad M_{P^R_{1, 1}, M_{1, 2}}
\] 
is non-trivial as otherwise $ \mathcal F^{(2)}_0$ would lift to a lattice VOA module and $P^R_{1, 1}$ would lift to a non-trivial triplet VOA module. But we know that neither of the two happens. 
The co-equalizer as described in section \ref{sec:liftlog} is here jus the co-kernel of the nilpotent part of $M_{P^R_{1, 1}, M_{1, 2}}\otimes M_{\mathcal F^{(2)}_0, \mathcal F_{\sqrt{-1}}}$. The image of the nilpotent part of $M_{P^R_{1, 1}, M_{1, 2}}\otimes M_{\mathcal F^{(2)}_0, \mathcal F_{\sqrt{-1}}}$ is clearly $P_{1, 2}^R\otimes \mathcal F_{\sqrt{-1}}$ and so the cokernel has Loewy diagram as in the following Figure \ref{fig:logprojbetagamma}.
 
\begin{figure}[tb]
\begin{center}
\begin{tikzpicture}[scale=0.70][thick,>=latex,
nom/.style={circle,draw=black!20,fill=black!20,inner sep=1pt}
]

\node(Left0) at  (8,0) [] {$X^R_{s,k}$:}; 
\node (Top1) at (15,2.5) [] {$M_{s, k}\otimes \mathcal F_{\lambda}$};
\node (Left1) at (11.5,0) [] {$M_{p-s, k+1}\otimes \mathcal F_{\lambda}$};
\node (Right1) at (18.5,0) [] {$M_{p-s, k-1}\otimes \mathcal F_{\lambda}$};
\node (Bot1) at (15,-2.5) [] {$M_{s, k}\otimes \mathcal F_{\lambda}$};
\draw [->] (Top1) -- (Left1);
\draw [->] (Top1) -- (Right1);
\draw [->] (Left1) -- (Bot1);
\draw [dashed, ->] (Top1) -- (Bot1);
\end{tikzpicture}
\captionbox{\label{fig:logprojbetagamma} New modules extending  the simple module $M_{s,k}\otimes \mathcal F_{\lambda}$ and that can lift to local $\mathcal B_2$-modules. One has $X^R_{s,k} \cong P^R_{s, k} \otimes \mathcal F_\lambda$ as $M(p)$-module and 
$X^R_{s,k} \cong M_{s, k} \otimes \mathcal F^{(2)}_\lambda  \oplus  M_{p-s, k+1} \otimes \mathcal F_\lambda \oplus  M_{p-s, k-1} \otimes \mathcal F_\lambda$ as Heisenberg VOA module. 
}{\rule{12cm}{0cm}}
\end{center}
\end{figure}

The computation of extensions is thus exactly the same as outlined in section \ref{sec:Clog} and thus we especially have that extensions of $\beta\gamma$-VOA are trivial:
\begin{equation}\label{eq:betagammaext}
\Ext_{\catlog}(\mathcal B_2, \mathcal B_2) = \mathbb C. 
\end{equation}

\subsection{Sub VOAs of many $\beta\gamma$ VOAs}

The construction here is somehow a generalization of the construction of the $\mathcal B_p$-algebras in \cite{CRW}. 
Consider now a sublattice $L\subset \mathbb Z^N=\alpha_1\mathbb Z \oplus \dots \oplus \alpha_n\mathbb Z$ (the product is given by $\alpha_i\alpha_j=\delta_{i, j}$) with orthogonal complement $L^\perp$ so that $L$ decomposes into cosets for the orthogonal sum:
\[
\mathbb Z^N = \bigoplus (L+\lambda) \oplus  (L^\perp \oplus \lambda^\perp).
\]
Here $\lambda$ and $\lambda^\perp$ are the coset representatives. Let $D$ be the diagonal isotropic sublattice of $\sqrt{-1}L\oplus L$.

Consider the lattice VOA $V_{\sqrt{-1}L} \otimes V_{\mathbb Z^N}$ which has $V_D\otimes V_{L^\perp}$ as sub VOA. Let $Q_1, \dots, Q_N$ be the screening charges such that the joint kernel of then on $V_{\mathbb Z^N}$ is just the VOA of $N$ $\beta\gamma$-VOAs. 
Define 
\[
B = \bigcap_{i=1}^N \text{ker}_{Q_i}(V_D\otimes V_{L^\perp}). 
\]
We would like to study extensions of the VOA $B$ in the category of logarithmic modules. For this, we need to know which modules lift from singlet times Heisemberg VOA modules to local modules of $B$. From the discussion of the $\beta\gamma$-VOA we see that this happens only if we can pair the nilpotent part of $H$ with a nilpotent action on Fock modules of the Heisenberg VOA. I.e. let $\catlog^{L}$ be the category of modules on which the endomorphism $\sum_{i=1}^N  a_i H_i$ acts semisimply if the element $\sum_{i=1}^N  a_i \alpha_i$ is in $L^\perp$. Then a necessary condition for a module
$M$ in $\catlog^{\boxtimes N}\boxtimes  (\catfocklog)^{\boxtimes N}$ to lift to a local $B$ module is that $M$ in $\catlog^{L}$.

The example we are interested is the root lattice $L=A_{N-1}$ with orthogonal complement $L^\perp = \sqrt{N}\mathbb Z$. In this case
$\catlog^{A_{N-1}}$ is the category $\catlog^N$ studied in section \ref{sec:manyClog}.

\subsubsection{$L_1(\mathfrak{psl}(N|N))$ and a diagonal simple current extension}\label{sec:pslNN}

The construction in this section follows very closely the one of the $\mathcal B_p$-algebras in \cite{CRW}. 
By Theorem 5.5 of \cite{CKLR} $L_1(\mathfrak{psl}(N|N))$ is a subquotient of a $U(1)$-orbifold of the superVOA of $N$-pairs of fermionic $bc$-ghost and bosonic $\beta\gamma$-ghosts. 
Consider now $N$ copies of the $p=2$ singlet together with the lattice superVOA of the lattice $\sqrt{-1}\mathbb Z^N$. The singlet VOA is the kernel of a screening charge $Q$ on the Heisenberg VOA, while the triplet is the kernel of the lattice VOA $V_{2\mathbb Z}$ and the symplectic fermions are the kernel of screenings of $V_{\mathbb Z}$. I.e. the inclusion of super VOAs
\[
\mathcal H \subset V_{2\mathbb Z}= \bigoplus_{k\in 2\mathbb Z} \mathcal F_k\subset V_{\mathbb Z}= \bigoplus_{k\in \mathbb Z} \mathcal F_k
\]
induces the inclusion of corresponding sub super VOAs
\[
M(2) = \text{ker}_Q(\mathcal H) \subset W(2) = \text{ker}_Q(\mathcal V_{2\mathbb Z})  \subset Fc = \text{ker}_Q(\mathcal V_{\mathbb Z}).
\]
Especially, the kernel of the screening on the Fock module $\mathcal F_k$ is the singlet simple current $M_{1, k+1}$.

Now, we can take the lattice $\sqrt{-1}\mathbb Z^N\oplus \mathbb Z^N$ and let $D$ be the diagonal isotropic sublattice. Then the kernel of all $N$ screenings $Q_1, \dots, Q_n$ restricted to this sublattice VOA is just $N$-pairs of $\beta\gamma$-ghost VOAs whose diagonal Heisenberg coset is $L_{-1}(\mathfrak{sl}_N)$ (except for $N=2$ where it is a rectangular W-algebra by Corollary 5.4 of  \cite{C2017}):
\[
\bigcap_{i=1}^N \text{ker}_{Q_i}(V_D) \cong \mathcal B_2^{\otimes N}\qquad, \qquad \text{Com}\left( \mathcal H, \mathcal B_2^{\otimes N}\right) \cong  \begin{cases} W^{\text{rect}}_{-5/2}(\mathfrak{sl}_4)  & \qquad N=2 \\ L_{-1}(\mathfrak{sl}_N) & \qquad \text{else} \end{cases}
\]

 Especially to every vector in $D$ we can associate the corresponding $(M(2) \otimes \mathcal H)^{\otimes N}$ -module. Now $\mathbb Z^N$ has the root lattice $A_{N-1}$ as sublattice and the orthogonal complement is $\sqrt{N}\mathbb Z$ and so $\mathbb Z$ decomposes as orthogonal sum
\[
\mathbb Z \cong \bigoplus_{n=0}^{N-1} (A_{N-1}+n\omega_1) \oplus \left(\sqrt{N} \mathbb Z + \frac{n}{\sqrt{N}}\right)
\]
with $\omega_1$ the first fundamental weight. Let $N\neq 2$ for the moment. Analogous decompositions hold for $\sqrt{-1}\mathbb Z$ and the diagonal isotro[ic sublattice $D$.
This means that we have the VOAs
\[
L_{-1}(\mathfrak{sl}_N) \cong \bigoplus_{\lambda \in A_{N-1}} M_\lambda
\]
with $M_\lambda$ the $M(2)^{\otimes N} \otimes \mathcal H^{\otimes (N-1)}$ associated to $\lambda$ and similarly the modules
\[
L_{-1}(\lambda) \cong  \bigoplus_{\mu \in \lambda +A_{N-1}} M_\mu
\]
for $\lambda$ in $P^+_1$. Then let
\[
X \cong \bigoplus_{\lambda \in P^+_1} L_{-1}(\lambda) \otimes L_{1}(\lambda)
\]
and $X\cong L_1(\mathfrak{psl}(N|N)$ for $N>2$.
The category of $N$-copies of $M(2)$ is a subcategory of the modules of $M(2)^{\otimes N} \otimes \mathcal H^{\otimes (N-1)}\otimes L_1(\mathfrak{sl}_N)$.
We are interested in the lifts of the modules $M_{\rho, K}$ with $\rho=(1, 1, \dots, 1)$ as before and $K$ an arbitry length $N$ vector with integer entries. Since all $M_{\rho, K}$ have the same conformal weight modulo $\mathbb Z$ all of them lift to local simple $X$ modules via induction. Inequivalent singlet modules induce to inequivalent local $X$-modules. Especially the Ext-algebra of $X$  is the same as the one of $N$-copies of the singlet, i.e. 
\begin{equation}
\begin{split}
\text{Ext}_\cat^\bullet(X, X) &\cong \text{Ext}_\cat^\bullet(M_{\rho, \rho}, M_{\rho, \rho})
\cong \bigoplus_{\substack{m=0 \\ m \ \text{even} }}^\infty \text{Sym}_{\frac{m}{2}}\mathbb C^N \cong \mathbb C[x_1^2, \dots, x_N^2]
\end{split}
\end{equation}
The algebra $A$ of \eqref{eq:algA} induces to a local object in a completion of the cateogory of local $X$-modules and thus gives rise to an extension 
\[
B = A \otimes_{\mathcal C} L_1(\mathfrak{psl}(N|N))
\]
of both VOAs. If $N=1$ these are just symplectic fermions, i.e. $L_1(\mathfrak{psl}(1|1)$, if $N=2$ then one can read of from Remark 9.11 of \cite{CG17} together with section 5 of \cite{C2017} that this is just $L_1(\mathfrak{psl}(2|2)$.

Singlet modules $M_{\rho, K_1}, M_{\rho, K_s}$ lift to the same extended VOA module if and only if 
\[
K_1 = K_2 \mod  \mathbb Z\rho, 
\]
and especially the Ext-algebra of $A$ and $B$ are the same 
\[
\text{Ext}_\cat^\bullet(B, B) \cong \text{Ext}_\cat^\bullet(A, A) \cong  \mathbb C[x_1^2, \dots, x_N^2, v_\pm^N]/ v_+^N  v_-^N - x_1^2 \cdot \dots \cdot x_N^2
\]
Chosing $\catlog$ instead of $\cat$ for singlet modules and also allowing Fock modules for self-extensions we get into the situation of sections \ref{sec:manyClog} and \ref{sec:manysinglets}. Those singlet modules that pair with Fock modules allow for extensions that lift to extensions of the VOAs $X$ and $B$. Hence the extended category is described by $\catlog^N$ of section \ref{sec:manyClog} and so especially 
\begin{equation}
\begin{split}
\text{Ext}_{\catlog}^\bullet(X, X) &\cong \text{Ext}_{\catlog^N}^\bullet(M_{\rho, \rho}, M_{\rho, \rho})
 \mathbb C[x^2]
\end{split}
\end{equation}
and 
\begin{equation}
\text{Ext}_{\catlog}^\bullet(B, B) \cong \text{Ext}_{\catlog^N}^\bullet(A, A) \cong  \mathbb C[x^2, v_\pm^N]/ v_+^N  v_-^N - x^{2N}.
\end{equation}
\\
\\
\noindent {\bf Acknowledgements.} T.~C. is supported by the Natural Sciences and Engineering Research Council of Canada (RES0020460).
K.C. and D.G. are supported by the NSERC
Discovery Grant program and by the Perimeter Institute for Theoretical
Physics. Research at Perimeter Institute is supported by the
Government of Canada through Industry Canada and by the Province of
Ontario through the Ministry of Research and Innovation. K.C. is grateful to Sasha Braverman for conversations about vertex algebras and Coulomb branch operators.
\appendix
\section{Computation of self-$\Ext$'s of Kac-Moody superalgebras}\label{app}

 In this section we study the vertex algebra which lives at the boundary of the Rozansky-Witten twist of a $3d$ $N=4$ gauge theory. We call this the Coulomb branch vertex algebra, even though the bulk operators that survive the Rozansky-Witten twist parameterize the Higgs branch.
 
The vertex algebra we will study is that associated to Dirichlet boundary conditions.  If the gauge group is $G$ and the matter consists of half-hypermultiplets in a  complex symplectic representation $V$, then the boundary vertex algebra is the WZW model for the super Lie algebra
\begin{equation} 
	\g_V = \g \oplus V[-1] \oplus \g^\vee [-2] 
\end{equation}
where the symbol $[-k]$ indicates a direct summand is placed in cohomological degree $k$.  (We only care about $\g_V$ as a $\bZ/2$ graded Lie algebra but this $\bZ$-grading is natural).  The Lie brackets are as follows: the bracket of $X \in \g$ with anything is given by the action of $\g$ on all the vector spaces appearing.  The bracket of two elements $v,w \in V[-1]$ with each other is given by
\begin{equation} 
	[v,w] = \tfrac{1}{2} \left(  \mu(v+w) - \mu(v) - \mu(w) \right) . 
\end{equation}
Here $\mu(v) \in \g^\vee$ is the moment map for the $G$ action on $V$, which is a $\g^\vee$-valued quadratic function on $V$.

Non-perturbatively, the WZW model for this super-algebra might be complicated to describe.  It will include the current algebra, as the perturbative sector, but also contributions from boundary monopoles.  These boundary monopoles can be described (abstractly) in terms of the Dolbeault homology of the affine Grassmannian for $G$ with coefficients in a super-vector bundle built from $V$ and $\g^\vee$.  

In this section, however, we will only perform a perturbative analysis, in which the vertex algebra is a current algebra that is easy to describe. It is generated by a super-current $J_a$ of spin one (where the index $a$ runs over a basis of $\g_V$).  The OPEs are the usual ones
\begin{equation} 
	J_a J_b \simeq \frac{1}{z} f^c_{ab} J_c + \omega_{ab} \frac{1}{z^2} 
\end{equation}
where $\omega_{ab}$ represents the natural invariant symmetric pairing on $\g_V$. This pairs $\g$ with $\g^\vee$ and $\Pi V$ with itself using the symplectic form. \footnote{There is a one-loop correction to this central charge, which takes the form of a standard 
central extension of $\g$. It is the difference of two terms, one proportional to the second Casimir of $V$ and one proportional to the second Casimir of $\g$. The correction does not affect the argument below. }

The Lie algebra $\g_V$ has a $\C^\times$ symmetry under which $V$ has weight $1$ and $\g_V$ has weight $2$, which scales the invariant pairing.  Because of this we can always normalize the level to be either one or zero.   The level zero algebra is the algebra of operators at the boundary when we twist the $N=4$ theory using a supercharge inside an $N=2$ subalgebra.  In this case, the bulk theory is not topological. We are therefore interested in the level one case.

Now let us compute the self-Ext's of the vacuum module. Modules for the vertex algebra are the same as representations of the universal enveloping algebra
\begin{equation} 
	U_{c = 1} ( \what{\g_V})  
\end{equation}
of the affine Lie algebra 
\begin{equation} 
	\what{\g_V} = \C \cdot c \oplus \g_V((z))
\end{equation} 
where we set the central parameter $c$ to be one.  We will compute the self-Ext's in the category of  modules for this (topological) associative algebra.

There is a sub-algebra 
\begin{equation} 
	U(\g_V[[z]]) \subset  U_{c = 1} ( \what{\g_V}) . 
\end{equation}
The vacuum module is
\begin{equation} 
	M = U_{c = 1} ( \what{\g_V}) \otimes_{  U(\g_V[[z]]) } \C.
\end{equation}
We can compute the derived endomorphisms of $M$ as follows:
\begin{align} 
	&	\op{RHom}_{ U_{c = 1} ( \what{\g_V})} (U_{c = 1} ( \what{\g_V}) \otimes_{  U(\g_V[[z]]) } \C,U_{c = 1} ( \what{\g_V}) \otimes_{  U(\g_V[[z]]) } \C)\\ 
	&=	
	\op{RHom}_{  U(\g_V[[z]]) } ( \C,U_{c = 1} ( \what{\g_V}) \otimes_{  U(\g_V[[z]]) } \C)  \\
	&= C^\ast( \g_V[[z]], M) 
\end{align}
where on the last line $C^\ast$ indicates Lie algebra cochains. We are using the standard fact that for any Lie algebra $\mf{l}$ and module $V$, we can identify $\op{RHom}_{U (\mf{l})}(\C, V)$ with $C^\ast(\mf{l}, V)$.  

We want to compare this with the algebra of bulk operators.  In the Rozansky-Witten twist, the algebra of bulk operators is the algebra of functions on the Higgs branch, which is the holomorphic symplectic reduction of $V$ by $G$.  Since we work in perturbation theory, we instead perform the reduction of $V$ by the action of the Lie algebra $\g$.  We perform this reduction in the derived sense.  First, we set the moment map to zero, by introducing fermionic variables $\eps_a$ living in $\g$ whose differentials are the components of the moment map: $\d \eps_a = \mu_a \in \Sym V^\vee$.  This gives us a differential graded algebra which, when we forget the differential, is the tensor product of $\wedge^\ast \g$ with $\Sym^\ast V^\vee$.  

Next, we take $\g$-invariants.  We again do this in the derived sense, by taking the Lie algebra cochains of $\g$ with coefficients in this differential graded algebra.  The result is a differential graded  algebra of the form $\wedge^\ast \g^\ast \otimes \wedge^\ast \g \otimes \Sym^\ast V^\vee$ whose differential has the form of the BRST operator in a $\b-\c$ ghost system.  One can identify this differential graded algebra with the Lie algebra cochains of $\g_V$.

The vacuum vector $\vac \in M$ is invariant under the action of $\g_V[[z]]$.  We therefore find cochain maps 
\begin{equation} 
C^\ast(\g_V) \to C^\ast(\g_V[[z]]) \to C^\ast(\g_V[[z]], M) . 
 \end{equation}
One can check that the cochain map
\begin{equation} 
C^\ast(\g_V) \to C^\ast(\g_V[[z]], M) \label{eqn_homomorphism} 
 \end{equation}
 is a homomorphism of associative algebras, where $C^\ast(\g_V[[z]], M)$ is given the algebra structure it acquires via the identification with $\op{RHom}_{U_{c =1}(\what{\g_V}) } (M, M)$.

Our task, therefore, is to show that the cochain map \ref{eqn_homomorphism} gives an isomorphism in cohomology.  To prove this we will use a spectral sequence.  The vacuum module $M$ has a filtration by sub-$\g_V[[z]]$ modules where $F^i M$ consists of those vectors which can be obtained from the vacuum vector by $\le i$ lowering operators.  The associated graded with respect to this filtration is 
\begin{equation} 
\op{Gr}^k  M = \Sym^k z^{-1} \g_V[z^{-1}].  
 \end{equation}
We therefore have a spectral sequence
\begin{equation} 
C^\ast(\g_V[[z]], \Sym^\ast (z^{-1} \g_V[z^{-1}]) \Longrightarrow  C^\ast(\g_V[[z]], M)  . 
 \end{equation}
On the left hand side, the $\g_V[[z]]$ module $ \Sym^\ast (z^{-1} \g_V[z^{-1}])$ is the symmetric algebra of the module $z^{-1} \g_V[z^{-1}]$, which can in turn be identified with the linear dual of $\g_V[[z]]$ (using the residue pairing).  We can view this symmetric algebra as being the exterior algebra of the dual of the fermionic vector space $\Pi \g_V[[z]]$.  Therefore we find
\begin{equation} 
C^\ast(\g_V[[z,\eps]] ) =  C^\ast(\g_V[[z]], \Sym^\ast (z^{-1} \g_V[z^{-1}]) ) 
 \end{equation}
where $\eps$ is a fermionic parameter.

We now need to compute the differential on the next page of the spectral sequence.  This differential involves those terms in the action of $\g_V[[z]]$ on $M$ which send
\begin{equation} 
\Sym^i z^{-1} \g_V[z^{-1}] \to \Sym^{i-1} z^{-1} \g_V[z^{-1}]. 
 \end{equation}
This term arises from the central extension, and is given by the pairing 
\begin{align} 
\g_V[[z]] \otimes z^{-1} \g_V[z^{-1}] & \to \C \\
X f(z) \otimes Y g(z) & \mapsto \omega(X,Y) \oint f \partial g. 
 \end{align}
In terms of the Lie algebra $C^\ast(\g_V[[z,\eps]])$ this term arises by turning it into a dg Lie algebra with differential $\eps \partial_z$.  

We find that the second page of the spectral sequence is given by Lie algebra cochains of this dg Lie algebra.   Now, the natural map 
\begin{equation} 
\g_V[[z,\eps]] \to \g_V
 \end{equation}
is a quasi-isomorphism of dg Lie algebras, where the left hand side is equipped with the differential $\eps \partial_z$.  It follows that the induced map on Lie algebra cochains 
\begin{equation} 
C^\ast(\g_V) \to C^\ast(\g_V[[z,\eps]])  
 \end{equation}
is also a quasi-isomorphism. We conclude that the map 
\begin{equation} 
C^\ast(\g_V) \to C^\ast(\g_V[[z]], M) 
 \end{equation}
is an isomorphism after passing to the cohomology of the second page of the spectral sequence. This is what we wanted to show.

\bibliographystyle{JHEP}

\bibliography{mono}

\providecommand{\href}[2]{#2}\begingroup\raggedright\begin{thebibliography}{10}

\bibitem{Nekrasov:2002qd}
N.~A. Nekrasov, {\it {Seiberg-Witten prepotential from instanton counting}},
  {\em Adv. Theor. Math. Phys.} {\bf 7} (2003), no.~5 831--864,
  [\href{http://arxiv.org/abs/hep-th/0206161}{{\tt hep-th/0206161}}].

\bibitem{Alday:2009aq}
L.~F. Alday, D.~Gaiotto, and Y.~Tachikawa, {\it {Liouville Correlation
  Functions from Four-dimensional Gauge Theories}},  {\em Lett. Math. Phys.}
  {\bf 91} (2010) 167--197, [\href{http://arxiv.org/abs/0906.3219}{{\tt
  arXiv:0906.3219}}].

\bibitem{Nekrasov:2010ka}
N.~Nekrasov and E.~Witten, {\it {The Omega Deformation, Branes, Integrability,
  and Liouville Theory}},  {\em JHEP} {\bf 09} (2010) 092,
  [\href{http://arxiv.org/abs/1002.0888}{{\tt arXiv:1002.0888}}].

\bibitem{Beem:2013sza}
C.~Beem, M.~Lemos, P.~Liendo, W.~Peelaers, L.~Rastelli, and B.~C. van Rees,
  {\it {Infinite Chiral Symmetry in Four Dimensions}},  {\em Commun. Math.
  Phys.} {\bf 336} (2015), no.~3 1359--1433,
  [\href{http://arxiv.org/abs/1312.5344}{{\tt arXiv:1312.5344}}].

\bibitem{Beem:2014rza}
C.~Beem, W.~Peelaers, L.~Rastelli, and B.~C. van Rees, {\it {Chiral algebras of
  class S}},  {\em JHEP} {\bf 05} (2015) 020,
  [\href{http://arxiv.org/abs/1408.6522}{{\tt arXiv:1408.6522}}].

\bibitem{Beem:2014kka}
C.~Beem, L.~Rastelli, and B.~C. van Rees, {\it {$ \mathcal{W} $ symmetry in six
  dimensions}},  {\em JHEP} {\bf 05} (2015) 017,
  [\href{http://arxiv.org/abs/1404.1079}{{\tt arXiv:1404.1079}}].

\bibitem{Bonetti:2018fqz}
F.~Bonetti, C.~Meneghelli, and L.~Rastelli, {\it {VOAs labelled by complex
  reflection groups and $4d$ SCFTs}},
  \href{http://arxiv.org/abs/1810.03612}{{\tt arXiv:1810.03612}}.

\bibitem{Gaiotto:2016wcv}
D.~Gaiotto, {\it {Twisted compactifications of 3d N = 4 theories and conformal
  blocks}},  \href{http://arxiv.org/abs/1611.01528}{{\tt arXiv:1611.01528}}.

\bibitem{Gaiotto:2017euk}
D.~Gaiotto and M.~Rap{\v c}{\'a}k, {\it {Vertex Algebras at the Corner}},
  \href{http://arxiv.org/abs/1703.00982}{{\tt arXiv:1703.00982}}.

\bibitem{CG17}
T.~Creutzig and D.~Gaiotto, {\it {Vertex Algebras for S-duality}},
  \href{http://arxiv.org/abs/1708.00875}{{\tt arXiv:1708.00875}}.

\bibitem{Costello:2018fnz}
K.~Costello and D.~Gaiotto, {\it {Vertex Operator Algebras and 3d $\mathcal
  N=4$ gauge theories}},  \href{http://arxiv.org/abs/1804.06460}{{\tt
  arXiv:1804.06460}}.

\bibitem{Seiberg:1996bs}
N.~Seiberg, {\it {IR dynamics on branes and space-time geometry}},  {\em Phys.
  Lett.} {\bf B384} (1996) 81--85,
  [\href{http://arxiv.org/abs/hep-th/9606017}{{\tt hep-th/9606017}}].

\bibitem{Seiberg:1996nz}
N.~Seiberg and E.~Witten, {\it {Gauge dynamics and compactification to
  three-dimensions}},  in {\em {The mathematical beauty of physics: A memorial
  volume for Claude Itzykson. Proceedings, Conference, Saclay, France, June
  5-7, 1996}}, pp.~333--366, 1996.
\newblock \href{http://arxiv.org/abs/hep-th/9607163}{{\tt hep-th/9607163}}.

\bibitem{deBoer:1996ck}
J.~de~Boer, K.~Hori, H.~Ooguri, Y.~Oz, and Z.~Yin, {\it {Mirror symmetry in
  three-dimensional theories, SL(2,Z) and D-brane moduli spaces}},  {\em Nucl.
  Phys.} {\bf B493} (1997) 148--176,
  [\href{http://arxiv.org/abs/hep-th/9612131}{{\tt hep-th/9612131}}].

\bibitem{Kapustin:1999ha}
A.~Kapustin and M.~J. Strassler, {\it {On mirror symmetry in three-dimensional
  Abelian gauge theories}},  {\em JHEP} {\bf 04} (1999) 021,
  [\href{http://arxiv.org/abs/hep-th/9902033}{{\tt hep-th/9902033}}].

\bibitem{Hanany:1996ie}
A.~Hanany and E.~Witten, {\it {Type IIB superstrings, BPS monopoles, and
  three-dimensional gauge dynamics}},  {\em Nucl. Phys.} {\bf B492} (1997)
  152--190, [\href{http://arxiv.org/abs/hep-th/9611230}{{\tt hep-th/9611230}}].

\bibitem{Gaiotto:2008ak}
D.~Gaiotto and E.~Witten, {\it {S-Duality of Boundary Conditions In N=4 Super
  Yang-Mills Theory}},  {\em Adv. Theor. Math. Phys.} {\bf 13} (2009), no.~3
  721--896, [\href{http://arxiv.org/abs/0807.3720}{{\tt arXiv:0807.3720}}].

\bibitem{Bullimore:2015lsa}
M.~Bullimore, T.~Dimofte, and D.~Gaiotto, {\it {The Coulomb Branch of 3d
  ${\mathcal{N}= 4}$ Theories}},  {\em Commun. Math. Phys.} {\bf 354} (2017),
  no.~2 671--751, [\href{http://arxiv.org/abs/1503.04817}{{\tt
  arXiv:1503.04817}}].

\bibitem{Nakajima:2015txa}
H.~Nakajima, {\it {Towards a mathematical definition of Coulomb branches of
  $3$-dimensional $\mathcal{N}=4$ gauge theories, I}},  {\em Adv. Theor. Math.
  Phys.} {\bf 20} (2016) 595--669, [\href{http://arxiv.org/abs/1503.03676}{{\tt
  arXiv:1503.03676}}].

\bibitem{Braverman:2016wma}
A.~Braverman, M.~Finkelberg, and H.~Nakajima, {\it {Towards a mathematical
  definition of Coulomb branches of $3$-dimensional $\mathcal N=4$ gauge
  theories, II}},  \href{http://arxiv.org/abs/1601.03586}{{\tt
  arXiv:1601.03586}}.

\bibitem{Yagi:2014toa}
J.~Yagi, {\it {$\Omega$-deformation and quantization}},  {\em JHEP} {\bf 08}
  (2014) 112, [\href{http://arxiv.org/abs/1405.6714}{{\tt arXiv:1405.6714}}].

\bibitem{Creutzig:2013hma}
T.~Creutzig and D.~Ridout, {\it {Logarithmic Conformal Field Theory: Beyond an
  Introduction}},  {\em J. Phys.} {\bf A46} (2013) 4006,
  [\href{http://arxiv.org/abs/1303.0847}{{\tt arXiv:1303.0847}}].

\bibitem{Kausch}
H.~G. Kausch, {\it {Symplectic fermions}},  {\em Nucl. Phys.} {\bf B583} (2000)
  513--541, [\href{http://arxiv.org/abs/hep-th/0003029}{{\tt hep-th/0003029}}].

\bibitem{Ridout:2008nh}
D.~Ridout, {\it {(hat)sl(2)(-1/2): A Case Study}},  {\em Nucl. Phys.} {\bf
  B814} (2009) 485--521, [\href{http://arxiv.org/abs/0810.3532}{{\tt
  arXiv:0810.3532}}].

\bibitem{Assel:2015oxa}
B.~Assel and J.~Gomis, {\it {Mirror Symmetry And Loop Operators}},  {\em JHEP}
  {\bf 11} (2015) 055, [\href{http://arxiv.org/abs/1506.01718}{{\tt
  arXiv:1506.01718}}].

\bibitem{CREUTZIG2019396}
T.~Creutzig and A.~R. Linshaw, {\it Cosets of affine vertex algebras inside
  larger structures},  {\em Journal of Algebra} {\bf 517} (2019) 396 -- 438,
  [\href{http://arxiv.org/abs/1407.8512}{{\tt arXiv:1407.8512}}].

\bibitem{Creutzig:2018ltv}
T.~Creutzig, D.~Gaiotto, and A.~R. Linshaw, {\it {S-duality for the large $N=4$
  superconformal algebra}},  \href{http://arxiv.org/abs/1804.09821}{{\tt
  arXiv:1804.09821}}.

\bibitem{Kausch:1990vg}
H.~G. Kausch, {\it {Extended conformal algebras generated by a multiplet of
  primary fields}},  {\em Phys. Lett.} {\bf B259} (1991) 448--455.

\bibitem{Adamovic:2007qs}
D.~Adamovic and A.~Milas, {\it {Logarithmic intertwining operators and
  W(2,2p-1)-algebras}},  {\em J. Math. Phys.} {\bf 48} (2007) 073503,
  [\href{http://arxiv.org/abs/math/0702081}{{\tt math/0702081}}].

\bibitem{Creutzig:2008an}
T.~Creutzig and P.~B. Ronne, {\it {The GL(1$|$1)-symplectic fermion
  correspondence}},  {\em Nucl. Phys.} {\bf B815} (2009) 95--124,
  [\href{http://arxiv.org/abs/0812.2835}{{\tt arXiv:0812.2835}}].

\bibitem{Witten:2005px}
E.~Witten, {\it {Two-dimensional models with (0,2) supersymmetry: Perturbative
  aspects}},  {\em Adv. Theor. Math. Phys.} {\bf 11} (2007), no.~1 1--63,
  [\href{http://arxiv.org/abs/hep-th/0504078}{{\tt hep-th/0504078}}].

\bibitem{Gorbounov:2016oia}
V.~Gorbounov, O.~Gwilliam, and B.~Williams, {\it {Chiral differential operators
  via Batalin-Vilkovisky quantization}},
  \href{http://arxiv.org/abs/1610.09657}{{\tt arXiv:1610.09657}}.

\bibitem{CG}
N.~Chriss and V.~Ginzburg, {\it Representation theory and complex geometry},
  {\em Birkh¨auser} (1997).

\bibitem{Gaiotto:2008sd}
D.~Gaiotto and E.~Witten, {\it {Janus Configurations, Chern-Simons Couplings,
  And The theta-Angle in N=4 Super Yang-Mills Theory}},  {\em JHEP} {\bf 06}
  (2010) 097, [\href{http://arxiv.org/abs/0804.2907}{{\tt arXiv:0804.2907}}].

\bibitem{AM1}
D.~Adamovic and A.~Milas, {\it {On the triplet vertex algebra W(p)}},  {\em
  Adv. Math.} {\bf 217} (2008) 2664--2699,
  [\href{http://arxiv.org/abs/0707.1857}{{\tt arXiv:0707.1857}}].

\bibitem{AM2}
D.~Adamovic and A.~Milas, {\it {The Structure of Zhu's algebras for certain
  W-algebras}},  \href{http://arxiv.org/abs/1006.5134}{{\tt arXiv:1006.5134}}.

\bibitem{Feiginetal}
B.~L. Feigin, A.~M. Gainutdinov, A.~M. Semikhatov, and I.~{\relax Yu}. Tipunin,
  {\it {Kazhdan-Lusztig correspondence for the representation category of the
  triplet W-algebra in logarithmic CFT}},  {\em Theor. Math. Phys.} {\bf 148}
  (2006) 1210--1235, [\href{http://arxiv.org/abs/math/0512621}{{\tt
  math/0512621}}]. [Teor. Mat. Fiz.148,398(2006)].

\bibitem{arakawa2016}
T.~Arakawa, {\it Rationality of admissible affine vertex algebras in the
  category ${\mathcal{o}}$},  {\em Duke Math. J.} {\bf 165} (01, 2016) 67--93.

\bibitem{AdaVer95}
D.~Adamovi\'{c} and A.~Milas, {\it Vertex operator algebras associated to
  modular invariant representations of {$A_1^{\left(1\right)}$}}, .
  {9509025}{q-alg}.

\bibitem{Creutzig:2013yca}
T.~Creutzig and D.~Ridout, {\it {Modular Data and Verlinde Formulae for
  Fractional Level WZW Models II}},  {\em Nucl. Phys.} {\bf B875} (2013)
  423--458, [\href{http://arxiv.org/abs/1306.4388}{{\tt arXiv:1306.4388}}].

\bibitem{Creutzig:2012sd}
T.~Creutzig and D.~Ridout, {\it {Modular Data and Verlinde Formulae for
  Fractional Level WZW Models I}},  {\em Nucl. Phys.} {\bf B865} (2012)
  83--114, [\href{http://arxiv.org/abs/1205.6513}{{\tt arXiv:1205.6513}}].

\bibitem{Feigin:2010xv}
B.~L. Feigin and I.~{\relax Yu}. Tipunin, {\it {Logarithmic CFTs connected with
  simple Lie algebras}},  \href{http://arxiv.org/abs/1002.5047}{{\tt
  arXiv:1002.5047}}.

\bibitem{Lentner:2017dkg}
S.~D. Lentner, {\it {Quantum groups and Nichols algebras acting on conformal
  field theories}},  \href{http://arxiv.org/abs/1702.06431}{{\tt
  arXiv:1702.06431}}.

\bibitem{Creutzig:2016uqu}
T.~Creutzig and A.~Milas, {\it {Higher rank partial and false theta functions
  and representation theory}},  {\em Adv. Math.} {\bf 314} (2017) 203--227,
  [\href{http://arxiv.org/abs/1607.08563}{{\tt arXiv:1607.08563}}].

\bibitem{Creutzig:2018lbc}
T.~Creutzig, {\it {Logarithmic W-algebras and Argyres-Douglas theories at
  higher rank}},  \href{http://arxiv.org/abs/1809.01725}{{\tt
  arXiv:1809.01725}}.

\bibitem{CGR}
T.~Creutzig, A.~M. Gainutdinov, and I.~Runkel, {\it {A quasi-Hopf algebra for
  the triplet vertex operator algebra}},
  \href{http://arxiv.org/abs/1712.07260}{{\tt arXiv:1712.07260}}.

\bibitem{HLZ1}
Y.-Z. Huang, J.~Lepowsky, and L.~Zhang, {\it {Logarithmic tensor category
  theory, I: Introduction and strongly graded algebras and their generalized
  modules}},  \href{http://arxiv.org/abs/1012.4193}{{\tt arXiv:1012.4193}}.

\bibitem{HLZ2}
Y.-Z. Huang, J.~Lepowsky, and L.~Zhang, {\it {Logarithmic tensor category
  theory, II: Logarithmic formal calculus and properties of logarithmic
  intertwining operators}},  \href{http://arxiv.org/abs/1012.4196}{{\tt
  arXiv:1012.4196}}.

\bibitem{HLZ3}
Y.-Z. Huang, J.~Lepowsky, and L.~Zhang, {\it {Logarithmic tensor category
  theory, III: Intertwining maps and tensor product bifunctors}},
  \href{http://arxiv.org/abs/1012.4197}{{\tt arXiv:1012.4197}}.

\bibitem{HLZ4}
Y.-Z. Huang, J.~Lepowsky, and L.~Zhang, {\it {Logarithmic tensor category
  theory, IV: Constructions of tensor product bifunctors and the compatibility
  conditions}},  \href{http://arxiv.org/abs/1012.4198}{{\tt arXiv:1012.4198}}.

\bibitem{HLZ5}
Y.-Z. Huang, J.~Lepowsky, and L.~Zhang, {\it {Logarithmic tensor category
  theory, V: Convergence condition for intertwining maps and the corresponding
  compatibility condition}},  \href{http://arxiv.org/abs/1012.4199}{{\tt
  arXiv:1012.4199}}.

\bibitem{HLZ6}
Y.-Z. Huang, J.~Lepowsky, and L.~Zhang, {\it {Logarithmic tensor category
  theory, VI: Expansion condition, associativity of logarithmic intertwining
  operators, and the associativity isomorphisms}},
  \href{http://arxiv.org/abs/1012.4202}{{\tt arXiv:1012.4202}}.

\bibitem{HLZ7}
Y.-Z. Huang, J.~Lepowsky, and L.~Zhang, {\it {Logarithmic tensor category
  theory, VII: Convergence and extension properties and applications to
  expansion for intertwining maps}},
  \href{http://arxiv.org/abs/1110.1929}{{\tt arXiv:1110.1929}}.

\bibitem{HLZ8}
Y.-Z. Huang, J.~Lepowsky, and L.~Zhang, {\it {Logarithmic tensor category
  theory, VIII: Braided tensor category structure on categories of generalized
  modules for a conformal vertex algebra}},
  \href{http://arxiv.org/abs/1110.1931}{{\tt arXiv:1110.1931}}.

\bibitem{HKL}
Y.-Z. Huang, A.~Kirillov, and J.~Lepowsky, {\it {Braided tensor categories and
  extensions of vertex operator algebras}},  {\em Commun. Math. Phys.} {\bf
  337} (2015), no.~3 1143--1159, [\href{http://arxiv.org/abs/1406.3420}{{\tt
  arXiv:1406.3420}}].

\bibitem{CKM}
T.~Creutzig, S.~Kanade, and R.~McRae, {\it {Tensor categories for vertex
  operator superalgebra extensions}},
  \href{http://arxiv.org/abs/1705.05017}{{\tt arXiv:1705.05017}}.

\bibitem{CKL}
T.~Creutzig, K.~Shashank, and A.~R. Linshaw, {\it {Simple current extensions
  beyond semi-simplicity}},  {\em Communications in Contemporary Mathematics}
  (2015) [\href{http://arxiv.org/abs/1511.08754}{{\tt arXiv:1511.08754}}].

\bibitem{CKLR}
T.~Creutzig, S.~Kanade, A.~R. Linshaw, and D.~Ridout, {\it {Schur-Weyl Duality
  for Heisenberg Cosets}},  {\em Transformation Groups} (Oct, 2018)
  [\href{http://arxiv.org/abs/1611.00305}{{\tt arXiv:1611.00305}}].

\bibitem{CRW}
T.~Creutzig, D.~Ridout, and S.~Wood, {\it {Coset Constructions of Logarithmic
  (1, p) Models}},  {\em Lett. Math. Phys.} {\bf 104} (2014) 553--583,
  [\href{http://arxiv.org/abs/1305.2665}{{\tt arXiv:1305.2665}}].

\bibitem{CM}
T.~Creutzig and A.~Milas, {\it {False Theta Functions and the Verlinde
  formula}},  {\em Adv. Math.} {\bf 262} (2014) 520--545,
  [\href{http://arxiv.org/abs/1309.6037}{{\tt arXiv:1309.6037}}].

\bibitem{CMR}
T.~Creutzig, A.~Milas, and M.~Rupert, {\it Logarithmic link invariants of
  $u_q^h(sl_2)$ and asymptotic dimensions of singlet vertex algebras},  {\em
  Journal of Pure and Applied Algebra} (2017) 3224--3247,
  [\href{http://arxiv.org/abs/1605.05634}{{\tt arXiv:1605.05634}}].

\bibitem{AR}
J.~Auger and M.~Rupert, {\it On infinite order simple current extensions of
  vertex operator algebras},  {\em Contemporary Mathematics} {\bf 711} (2018)
  [\href{http://arxiv.org/abs/arXiv:1711.05343}{{\tt arXiv:1711.05343}}].

\bibitem{C2017}
T.~Creutzig, {\it W-algebras for argyres--douglas theories},  {\em European
  Journal of Mathematics} {\bf 3} (Sep, 2017) 659--690,
  [\href{http://arxiv.org/abs/1701.05926}{{\tt arXiv:1701.05926}}].

\end{thebibliography}\endgroup

\end{document}